
\def\bold#1{\setbox0=\hbox{$ #1$}%
     \kern-.010em\copy0\kern-\wd0
     \kern.030em\copy0\kern-\wd0
     \kern-.010em\raise.0033em\box0 }

\def\ls#1{\ifmath{_{\lower1.5pt\hbox{$\scriptstyle #1$}}}}
\def\ifmath#1{\relax\ifmmode #1\else $#1$\fi}
\def\SCIPP{\centerline {\it Santa Cruz Institute for Particle Physics}
  \centerline{\it University of California, Santa Cruz, CA 95064}}
\def\abs#1{\left|#1\right|}
\def\frac#1#2{{#1\over#2}}
 \def\sqr#1#2{{\vcenter{\hrule height.#2pt                       %
  \hbox{\vrule width.#2pt height#1pt \kern#1pt \vrule width.#2pt}%
   \hrule height.#2pt}}}                                         %
 \def\square{\mathchoice\sqr74\sqr74\sqr74\sqr74}                %
 \def\dalam{\lower-.7pt\hbox{$\square$}}


\def\smgaugegroup{{\rm SU(3)_c\times
SU(2)_L\times U(1)_Y}}
\def\sutwouone{{\rm SU(2)_L\times U(1)_Y}}

\def\uone{{\rm U(1)}}
\def\uy{{\rm U(1)_Y}}
\def\ug{{\rm U(1)_g}}
\def\sutwol{{\rm SU(2)_L}}

\def\suc{{\rm SU(3)_c}}
\def\uem{{\rm U(1)\ls{\rm EM}}}


\def\d{{\rm d}}
\def\diag{{\rm diag}}

\def\det{{\rm det}}

\def\hc{{\rm  h.c.}}



\def\calo{{\cal O}}

\def\calm{{\cal M}}

\def\calv{{\cal V}}
\def\call{{\cal L}}


\def\hl{{h^0}}
\def\ha{{A^0}}
\def\hh{{H^0}}
\def\hpm{{H^{\pm}}}


\def\mtt{m_t^2}
\def\mt{m_t}
\def\mbb{m_b^2}
\def\mb{m_b}
\def\mweak{M_{\rm weak}}
\def\mweakk{\mweak^2}
\def\mhl{m_{h^0}}
\def\mhh{m_{H^0}}

\def\mhpm{m_{H^{\pm}}}
\def\mha{m_{A^0}}
\def\mw{m_W}
\def\mww{m_W^2}
\def\mz{m_Z}
\def\mzz{m_Z^2}

\def\msusy{M_{\rm SUSY}}
\def\msusyy{M_{\rm SUSY}^2}


\def\tb  {t_{\beta}}
\def\tbb  {\tb^{-1}}
\def\sw  {s_W}
\def\cw  {c_W}
\def\sb  {s_{\beta}}
\def\cb  {c_{\beta}}

\def\ctwob  {c_{2\beta}}
\def\sa  {s_{\alpha}}
\def\ca  {c_{\alpha}}
\def\sab  {s_{\alpha+\beta}}
\def\cab  {c_{\alpha+\beta}}
\def\sba  {s_{\beta-\alpha}}
\def\cba  {c_{\beta-\alpha}}

\def\tanb{\tan\beta}
\def\sinb{\sin\beta}
\def\cosb{\cos\beta}

\def\sinbma{\sin(\beta-\alpha)}
\def\cosbma{\cos(\beta-\alpha)}
\def\cww{c_W^2}
\def\sww{s_W^2}


\def\sqr#1#2{{\vcenter{\vbox{\hrule height.#2pt
         \hbox{\vrule width.#2pt height#1pt \kern#1pt
            \vrule width.#2pt}
          \hrule height.#2pt}}}}


\def\one{{\bold 1}}


\def\fivethirds{\ifmath{{\textstyle{5\over 3}}}}
\def\half{\ifmath{{\textstyle{1 \over 2}}}}
\def\fivehalf{\ifmath{{\textstyle{5 \over 2}}}}
\def\fivetwelfth{\ifmath{{\textstyle{5 \over 12}}}}

\def\fourthirds{\ifmath{{\textstyle{4 \over 3}}}}
\def\third{\ifmath{{\textstyle{1 \over 3}}}}
\def\twothirds{\ifmath{{\textstyle{2\over 3}}}}
\def\fourth{\ifmath{{\textstyle{1\over 4}}}}

\def\fifteenfourth{\ifmath{{\textstyle{15\over 4}}}}

\def\eighth{\ifmath{{\textstyle{1 \over 8}}}}
\def\threeighth{\ifmath{{\textstyle{3 \over 8}}}}

\def\ninehalf{\ifmath{{\textstyle{9 \over 2}}}}
\def\seventeentwelfth{\ifmath{{\textstyle{17 \over12}}}}
\def\ninefourth{\ifmath{{\textstyle{9 \over 4}}}}
\def\threehalf{\ifmath{{\textstyle{3 \over 2}}}}

\overfullrule0pt
\Pubnum={SCIPP 91/33}
\date={July 1992}
\pubtype{Revised: \cr
         June 1993}
\titlepage
\vbox to 2cm{}
\title{{\bf The Renormalization-Group Improved Higgs Sector of
the Minimal Supersymmetric Model}
\foot{Work supported in part by the U.S.~Department of Energy.}}
\author{Howard E. Haber and Ralf Hempfling}
\vskip .1in
\SCIPP
\vskip .2in
\singlespace
\vfill
\centerline{\bf Abstract}

In the minimal supersymmetric model (MSSM) all Higgs self-coupling
parameters are related to gauge couplings at tree-level.
Leading-logarithmic radiative corrections to these quantities can
be summed using renormalization group techniques.   By this procedure
we obtain complete leading-log radiative corrections to the Higgs masses,
the CP-even Higgs mixing angle, and trilinear Higgs couplings.
Additional corrections due to squark mixing can be explicitly
incorporated into this formalism. These
results incorporate nearly all potentially large corrections.  Mass
shifts to the neutral CP-even
Higgs bosons grow with the fourth power of the
top-quark mass and can be significant.  The
phenomenological consequences of these results are examined.
\vskip .1in
\vfill
\submit{Physical Review D}
\vfill
\endpage

\chapter{Introduction}
Supersymmetry is one of the most promising theoretical ideas
that attempts to
\REF\susyrev{H.P. Nilles, {\sl Phys.~Rep.} {\bf 110}, 1 (1984);
H.E. Haber and G.L. Kane, {\sl Phys.~Rep.} {\bf 117}, 75 (1985);
R. Barbieri, {\sl Riv. Nuovo Cimento} {\bf 11}, 1 (1988).}
explain the
origin of the scale of electroweak interactions.
The minimal supersymmetric extension of the Standard Model (MSSM)
is the most economical among models of
this type\refmark\susyrev, and deserves
close examination as a candidate for a model of physics beyond
the Standard Model (SM).
In the MSSM, one simply adds a supersymmetric
partner to every quark, lepton and gauge boson.
In addition, the MSSM must possess two Higgs doublets in
order to give masses to up and down type fermions in a manner consistent
with supersymmetry (and to avoid gauge anomalies introduced by
the fermionic superpartners of the Higgs bosons).
\REF\hhg{J.F. Gunion, H.E. Haber, G.L. Kane, and S.
Dawson, \it The Higgs  Hunter's Guide, \rm (Addison-Wesley, Redwood City,
CA, 1990).}
The Higgs sector of the MSSM is greatly
constrained by supersymmetry\refmark\hhg.
All quartic Higgs coupling constants are related
to electroweak gauge coupling constants, which
imposes various restrictions on the tree-level Higgs
masses and couplings.
In particular, all tree-level
Higgs parameters can be expressed in terms of one physical Higgs mass
and the ratio of vacuum
expectation values, $\tanb\equiv v_2/v_1$.

\REF\gira{L. Girardello and M.T. Grisaru,
{\sl Nucl. Phys.} {\bf B194}, 65 (1982).}
\REF\berger{M.S. Berger, {\sl Phys. Rev.} {\bf D41}, 225 (1990).}
\REF\leter{H.E. Haber and R. Hempfling, {\sl Phys. Rev. Lett.} {\bf 66},
1815 (1991).}
\REF\rest{Y. Okada, M. Yamaguchi and T. Yanagida,
{\sl Prog. Theor. Phys.} {\bf 85}, 1 (1991);
J. Ellis, G. Ridolfi and F. Zwirner, {\sl Phys. Lett.}
{\bf B257}, 83 (1991).}
\REF\barbi{R. Barbieri, M. Frigeni and F. Caravaglios
{\sl Phys. Lett.} {\bf B258}, 167 (1991).}
\REF\yama{Y. Okada,
M. Yamaguchi and T. Yanagida, {\sl Phys.
Lett.} {\bf B262}, 54 (1991).}
\REF\erz{J. Ellis, G. Ridolfi and F. Zwirner, {\sl Phys. Lett.}
{\bf B262}, 477 (1991).}
\REF\quiros{J.R. Espinosa and M. Quir\'os,
{\sl Phys. Lett.} {\bf B266}, 389 (1991).}
\REF\yam{A. Yamada, {\sl Phys. Lett.} {\bf B263}, 233 (1991).}
\REF\ellis{A. Brignole, J. Ellis, G. Rudolfi and F. Zwirner,
\sl Phys. Lett. \bf B271, \rm 123 (1991)
[E: \bf B273, \rm 550 (1991)].}
\REF\hiroshima{H.E. Haber, in {\it
Proceedings of the International Workshop on Electroweak Symmetry
Breaking}, Hiroshima, Japan, 12-15 November 1991, edited by
W.A. Bardeen, J. Kodaira and T. Muta (World Scientific,
Singapore, 1992), p.~225.}
\REF\pokor{P.H. Chankowski, S. Pokorski and J. Rosiek, {\sl Phys. Lett.}
{\bf B274}, 191 (1992); {\bf B286}, 307 (1992).}
\REF\diaz{
M.A. D\'\i az and H.E. Haber, {\sl Phys. Rev.} {\bf D46}, 3086 (1992).}
\REF\brig{A. Brignole, {\sl Phys. Lett.} {\bf B281}, 284 (1992).}
\REF\berkeley{D.M. Pierce, A. Papadopoulos and S. Johnson,
\sl Phys. Rev. Lett. \bf 68\rm, 3678 (1992).}
\REF\drees{M. Drees and M.M. Nojiri, {\sl Phys. Rev.} {\bf D45}, 2482
(1992).}
\REF\sasaki{K. Sasaki, M.
Carena and C.E.M. Wagner, {\sl Nucl. Phys.} {\bf B381}, 66 (1992);
P.H. Chankowski, S. Pokorski and J. Rosiek, {\sl Phys. Lett.}
{\bf B281}, 100 (1992).}
\REF\bbsp{V. Barger, M.S. Berger, A.L. Stange and R.J.N. Phillips,
\sl Phys. Rev. \bf D45\rm, 4128 (1992).}%
Any realistic supersymmetric model must incorporate supersymmetry
breaking in the low-energy theory.  This breaking is parametrized by
adding soft supersymmetry breaking mass terms for the
squarks, sleptons and gauginos, and trilinear
Higgs-squark-squark and Higgs-slepton-slepton interactions
which are proportional to the so-called $A$ parameters\refmark\gira.
Due to these supersymmetry breaking terms, one finds that the
tree-level relations among Higgs masses and couplings acquire
radiative corrections.
It has been shown that these corrections can indeed become
very substantial if the top quark mass is much larger than
$\mz$\refmark{\berger-\bbsp}.
For example, the tree-level bound, $\mhl\leq\mz$ receives a radiative
correction of order $g_2^2\mt^4/\mzz\ln(M_{\tilde t}^2/\mtt)$
which raises the upper limit of $\mhl$ by as much as 20~(50)~GeV for a
top-quark mass of $\mt = 150$ (200)~GeV\refmark\leter.  The
logarithmic enhancement factor is a remnant of the cancellation
of divergences generated by virtual particles (in this case the
top quark and its supersymmetric partners).  Similar logarithmic
corrections
also arise when contributions from other sectors of the theory
are incorporated.

While the exact one-loop radiative
corrections to mass sum rules and Higgs mass bounds can be obtained
in a straightforward manner, the radiative corrections
to individual CP-even Higgs masses and Higgs interactions are far more
complex.  For instance,
computations of the latter type require a careful definition of
the parameter $\tanb$.
However, significant simplification can be achieved if we include only
the leading logarithmic radiative corrections.
In this approximation, the definitions of $\tanb$ and the
CP-even Higgs mixing angle
$\alpha$ are unambiguous and can easily be related to
physical observables.
The goal of this paper is to
construct a low-energy effective Lagrangian from which one can directly
obtain the leading contributions to the radiatively corrected
Higgs masses and couplings.

One possible procedure for
achieving this goal is to compute the terms of the
full one-loop effective action that depend on the scalar Higgs fields.
This requires the computation of the effective potential, $V_{\rm eff}$,
the coefficient of the scalar kinetic energy term, $Z_{\rm eff}$,
and higher derivative terms.  Let us assume that the supersymmetry
breaking scale ($\msusy$) is somewhat higher than the electroweak
breaking scale.%
\foot{If $\msusy$ is roughly equal to
the scale of electroweak symmetry breaking, then the size of the
radiative corrections discussed in this paper will be rather small.
In this case, one will need exact one-loop computations to
determine reliably the effects of the radiative corrections.}
Here, we use $\msusy$ to denote the mass scale that characterizes
supersymmetry breaking.  (For now, we ignore the possibility
of multiple supersymmetric thresholds,
which will be addressed later in this paper.)  If one expands
the effective action about the Higgs vacuum expectation value and
discards all terms of order $\mz^2/\msusy^2$, then it suffices to keep
only those terms of dimension 4 or less.
Note that at this stage, it is not strictly correct to use the effective
potential to compute the one-loop Higgs masses and couplings. One must
first rescale the scalar fields (\ie, wave function renormalization)
in order that the scalar kinetic
energy terms are canonical.  The end result is a scalar potential
that is polynomial in the scalar fields (with terms of dimension 4 or
less), whose coefficients reflect the one-loop radiative corrections.
The leading one-loop correction terms will depend logarithmically on
$\msusy$, which suggests that one can use
renormalization group methods to explicitly
identify the leading logarithmic terms\refmark{\barbi,\yama,\hiroshima,%
\sasaki}.
In this paper, we shall employ the renormalization group method
for two reasons: (i) simplicity, and (ii)
the integration of the one-loop renormalization group equations (RGEs)
effectively sums the leading logarithmic contributions to all orders
in perturbation theory.
This method can be extended to include the effects of supersymmetric
thresholds.  In this paper,
we derive RGEs that incorporate terms which are logarithmic
in the ratio of threshold masses to one-loop.  However, to sum such
effects to all orders significantly  complicates the analysis
and is beyond the scope of this paper.

In addition to neatly summarizing the most important radiative
corrections, the direct identification of the terms logarithmic in
$\msusy$ provides an important check for a more precise (and hence
more complicated) explicit one-loop computation.
Moreover, the full RGE analysis
yields leading log terms to all orders in perturbation theory,
and hence provides some
information on the contributions that lie beyond the one-loop
approximation\refmark\quiros.
This provides a check on the reliability of the
one-loop results.  However, it is important to realize
that the renormalization group technique may not detect all significant
terms of the one-loop radiative corrections.  For example, if the
$A$-terms that control squark mixing are large, then radiative correction
terms of order $A^2/\msusy^2$ may be enhanced by powers of the top-quark
mass\refmark{\erz,\ellis,\berkeley,\drees,\bbsp}.
Such terms can then compete with (and may be more important than) the
leading log terms identified above.  Thus, an important goal of
this paper is to demonstrate how to include such terms within the
renormalization group approach in a consistent manner.  The end result
of our work
is a simple and powerful technique that includes in a transparent
fashion the most important radiative corrections to the MSSM Higgs
sector.

In section 2,
we first examine the
general non-supersymmetric two Higgs doublet model at tree-level.
The most general two-Higgs-doublet model potential depends on
three mass parameters and seven dimensionless couplings. The Higgs mass
matrices and three-point Higgs vertices are obtained in terms of these
parameters. In section 3, we specialize to the MSSM. Radiative
corrections to the Higgs sector parameters are obtained by
renormalization group evolution.
Supersymmetry implies definite relations among the Higgs self-couplings
in terms of the gauge couplings. These relations are used as
boundary conditions for the renormalization group equations at a scale
$\msusy$ which characterizes the scale of supersymmetric particle masses.
The Higgs couplings are then evolved according to renormalization group
equations (RGEs) and the radiatively corrected Higgs masses and couplings
are computed in terms of parameters at the electroweak scale.
In the evolution of couplings, we first assume that the effective
low-energy theory at the electroweak scale contains two
light Higgs doublets. Another possibility is to assume that
the second Higgs doublet is significantly heavier than the electroweak
scale. In this case the low-energy effective theory is
identical to the SM with one physical Higgs boson. This case is discussed
section 4. In our analysis, an important parameter is $\tanb$,
the ratio of vacuum expectation values. The relation of $\tanb$ to
physically measurable observables is elucidated in section 5.

The analysis described above is equivalent to summing the leading
log
radiative corrections to all orders in perturbation theory.
This approximation summarizes almost all potentially large radiative
corrections. However,
if squark mixing effects are substantial, an additional set
of non-logarithmic corrections may be important.
We incorporate these terms into our analysis in section 6. Numerical
results and their phenomenological implications are given in section 7.
The importance of using RGE-improved results for the Higgs masses
is subject of
section 8. Final conclusions are given in section 9.
Details of the RGE analysis and the incorporation of important
non-leading
log terms are relegated to four appendices.
\REF\gbhs{J.F. Gunion, R. Bork, H.E. Haber and A. Seiden,
{\sl Phys. Rev.} {\bf D46}, 2040 (1992).}
\REF\ghk{J.F. Gunion, H.E. Haber and C. Kao,
\sl Phys Rev. \bf D46\rm , 2907 (1992).}
\REF\hhn{H.E. Haber, R. Hempfling and Y. Nir,
{\sl Phys. Rev.} {\bf D46}, 3015 (1992).}
\REF\ralfthesis{R. Hempfling, Ph.D. dissertation,
University of California, Santa Cruz, SCIPP 92/28 (1992).}

Some results of this paper were first described in ref.~[\hiroshima]
and used in the phenomenological analyses presented in
refs.~[\gbhs--\hhn].  Additional details can be found in
ref.~[\ralfthesis].

\chapter{The General Two-Higgs-Doublet Model}
We begin with a brief review of the general (non-supersymmetric)
two-Higgs doublet extension of the Standard Model\refmark\hhg.
Let $\Phi_1$ and $\Phi_2$ denote two complex $Y=1$, SU(2)$_L$ doublet
scalar fields.  We introduce the notation
$$\Phi_n =
\left(\matrix{H_n^+\cr (H_n^0+iA_n^0)/\sqrt2}\right)\,. \eqn\defphi$$
The most general gauge invariant scalar potential is given by
$$\eqalign{\calv&=m_{11}^2\Phi_1^\dagger\Phi_1+
m_{22}^2\Phi_2^\dagger\Phi_2 -[m_{12}^2\Phi_1^\dagger\Phi_2+\hc]\cr
&~~+\half\lambda_1(\Phi_1^\dagger\Phi_1)^2
+\half\lambda_2(\Phi_2^\dagger\Phi_2)^2
+\lambda_3(\Phi_1^\dagger\Phi_1)(\Phi_2^\dagger\Phi_2)
+\lambda_4(\Phi_1^\dagger\Phi_2)(\Phi_2^\dagger\Phi_1)\crr
&~~+\left\{\half\lambda_5(\Phi_1^\dagger\Phi_2)^2
+\big[\lambda_6(\Phi_1^\dagger\Phi_1)
+\lambda_7(\Phi_2^\dagger\Phi_2)\big]
\Phi_1^\dagger\Phi_2+\hc\right\}\,.\cr}\eqn\pot$$
In most discussions of two-Higgs-doublet models, the terms proportional
to $\lambda_6$ and $\lambda_7$ are absent.  This can be achieved by
imposing a discrete symmetry $\Phi_1\to -\Phi_1$ on the model.  Such a
symmetry would also require $m_{12}=0$ unless we allow a
soft violation of this discrete symmetry by dimension-two terms.\foot{%
This latter requirement is sufficient to guarantee the absence of
Higgs-mediated tree-level flavor changing neutral currents.}
For the moment, we will refrain from setting any of the coefficients
in eq.~\pot\ to zero.  In principle, $m_{12}^2$, $\lambda_5$,
$\lambda_6$ and $\lambda_7$ can be complex.  In this paper, we shall
ignore the possibility of CP-violating effects in the Higgs sector
by choosing all coefficients in eq.~\pot\ to be real.
The fields will
develop non-zero vacuum expectation values (VEVs) if the mass matrix
$m_{ij}^2$ has at least one negative eigenvalue. Imposing CP invariance
and U(1)$\ls{\rm EM}$ gauge symmetry, the minimum of the potential is
$$\langle \Phi_1 \rangle={1\over\sqrt{2}}
\pmatrix{0\cr v_1\cr}, \qquad \langle \Phi_2\rangle=
{1\over\sqrt{2}}\pmatrix{0\cr v_2\cr}\,,\eqn\potmin$$
where the $v_i$ can be chosen to be real.  The VEVs have been normalized
so that $\mww = \fourth g_2^2(v_1^2+v_2^2)$.
It is convenient to introduce
the following notation:
$$v^2\equiv v_1^2+v_2^2\,,
\qquad\qquad\tb\equiv\tanb\equiv{v_2/v_1}\,.\eqn\polar$$
The gauge symmetry will then be broken
spontaneously. As a result three of the eight degrees of freedom of the
original Higgs doublets are eaten by the $W^\pm$ and $Z$. The remaining
five physical Higgs particles are: two CP-even scalars ($\hh$ and
$\hl$, with $\mhl\leq \mhh$), one CP-odd scalar ($\ha$) and a charged
Higgs pair ($\hpm$). The mass parameters $m_{11}$ and $m_{22}$ can be
eliminated by imposing the minimization conditions
$$\eqalign{&m_{11}^2 - \tb m_{12}^2
+\half v^2\cb^2\left(\lambda_1
+3\lambda_6\tb+\widetilde\lambda_3\tb^2
+\lambda_7\tb^3\right)=0\,,\cr
&m_{22}^2 - \tbb m_{12}^2 +\half
v^2\sb^2\left(\lambda_2 +3\lambda_7\tbb+\widetilde
\lambda_3\tb^{-2}+\lambda_6\tb^{-3} \right)=0\,,
}\eqn\mincon$$
where
we have introduced the following abbreviations: $\sb\equiv\sinb$ and
$\cb\equiv\cosb$ and
$$
\widetilde\lambda_3\equiv\lambda_3+\lambda_4+\lambda_5\,.\eqn\ltilde
$$
It then follows that the mass matrices of the CP-even, CP-odd and the
charged scalars are given by
$$
\eqalign{
\calm_{A^0}^2&= \left[m_{12}^2-\half
v^2(2\lambda_5\sb\cb+\lambda_6\cb^2+\lambda_7\sb^2)\right]
\left(\matrix{\tb&-1\cr -1&\tbb}\right)\,,\crr
\calm_{H^{\pm}}^2&= \left[m_{12}^2-\half
v^2(\lambda_4\sb\cb+\lambda_5\sb\cb+\lambda_6\cb^2+
\lambda_7\sb^2)\right] \left(\matrix{\tb&-1\crr
-1&\tbb}\right)\,,\cr
\calm_{H^0}^2&= m_{12}^2
\left(\matrix{\tb&-1\cr -1&\tbb}\right)\crr
&~~~+\half v^2\sb\cb
\left(\matrix{2\lambda_1\tbb+3\lambda_6- \lambda_7\tb^2
&2\widetilde\lambda_3+3(\lambda_6 \tbb+\lambda_7\tb)\cr
2\widetilde\lambda_3+3(\lambda_6 \tbb+\lambda_7\tb)
&2\lambda_2\tb+3\lambda_7-\lambda_6\tb^{-2}}
\right)\,.\cr}\eqn\massma$$
The first two mass matrices
possess a zero eigenvalue corresponding to the
Goldstone bosons ($G^0,G^{\pm}$). The masses of the
physical Higgs particles are given by
$$\eqalign{
\mha^2&=\tr(\calm_{A^0}^2)=
{m_{12}^2\over \sb\cb}-\half
v^2\big(2\lambda_5+\lambda_6\tbb+\lambda_7\tb\big)\,,\cr
m_{H^{\pm}}^2&=\tr(\calm_{H^{\pm}}^2)= {m_{12}^2\over
\sb\cb}-\half
v^2\big(\lambda_4+\lambda_5+\lambda_6\tbb+\lambda_7\tb\big)\,.
}\eqn\mamthree$$
If we now substitute the remaining mass parameter $m_{12}^2$ in favor
of $\mha^2$ we find  the following expressions for the charged Higgs
mass and the neutral CP-even Higgs boson mass matrix
$$\eqalignno{%
m_{H^{\pm}}^2 &= m_{A^0}^2 + \half
v^2(\lambda_5-\lambda_4)\,,  &\eqnalign\dchrgd
\cr\crr
\calm^2_{H^0} &= m_{A^0}^2\left(\matrix{\sb^2&-\sb\cb\cr
-\sb\cb&\cb^2}\right)   &\crr
+&v^2 \left(\matrix{\lambda_1\cb^2+2\lambda_6\sb\cb+\lambda_5\sb^2
 &(\lambda_3+\lambda_4)\sb\cb+\lambda_6 \cb^2+\lambda_7\sb^2\cr
(\lambda_3+\lambda_4)\sb\cb+\lambda_6 \cb^2+\lambda_7\sb^2&
\lambda_2\sb^2+2\lambda_7\sb\cb+\lambda_5\cb^2\cr}\right). \qquad
&\eqnalign\massmhh}$$
The CP-even Higgs mass eigenvalues
are given by
$$m^2_{\hh,\hl}=\half\biggl\{\tr\calm^2\pm
\sqrt{[\tr\calm^2]^2-4\,\det\,\calm^2}\biggl\}\,,\eqn\massev$$
where $\calm^2\equiv\calm_{\hh}^2$ and the mixing angle $\alpha$ is
obtained from $$\eqalign{&\sin 2\alpha={2\calm^2_{12}\over
\sqrt{[\tr\calm^2]^2-4\,\det\,\calm^2}}\,,\cr &\cos
2\alpha={\calm^2_{11}-\calm^2_{22}\over
\sqrt{[\tr\calm^2]^2-4\,\det\,\calm^2}}\,.\cr}\eqn\defalpha$$

The phenomenology of the two-Higgs doublet model depends in detail on
the various couplings of the Higgs bosons to gauge bosons, Higgs
bosons and fermions.  The Higgs couplings to gauge bosons follow from
gauge invariance and are thus model independent.
Most of these
couplings are proportional to either $\sin(\beta-\alpha)$ or
$\cos(\beta-\alpha)$.
In contrast, the Higgs couplings
to fermions are model dependent, although their form is often
constrained by discrete symmetries that are imposed in order to
avoid tree-level flavor changing neutral currents mediated by Higgs
\REF\gw{S. Glashow and S. Weinberg, {\sl Phys. Rev.} {\bf D15}
1958 (1977).}
exchange\refmark\gw.
An example of a model that respects this constraint is one in
which one Higgs doublet (before symmetry
breaking) couples exclusively to down-type fermions and the other
Higgs doublet couples exclusively to up-type
fermions.  This is
the pattern of couplings found in the minimal supersymmetric model
(MSSM).  Detailed Feynman rules can be found in ref.~[\hhg].
Finally,
the 3-point and 4-point Higgs self-couplings depend on the
two-Higgs-doublet potential [eq.~\pot].  The Feynman rules for
the most important trilinear Higgs vertices are listed below:
$$\eqalign{%
&g\ls{h^0A^0A^0} =
   {2\mw\over g_2}\bigl[ \lambda_1\sb^2\cb\sa
   -\lambda_2\cb^2\sb\ca -\widetilde\lambda_3
   (\sb^3\ca-\cb^3\sa) +2\lambda_5\sba\cr
&\qquad\qquad\qquad -\lambda_6\sb\big(\cb\sab+\sa
c_{2\beta}\big)-\lambda_7\cb\big(\ca c_{2\beta}-\sb\sab\big)\bigr]\,,\crr
&g\ls{H^0A^0A^0} = {-2\mw\over g_2}\bigl[
   \lambda_1\sb^2\cb\ca+\lambda_2\cb^2\sb\sa+\widetilde\lambda_3
   (\sb^3\sa+\cb^3\ca) -2\lambda_5\cba\cr
&\qquad\qquad\qquad -\lambda_6\sb\big(\cb\cab+\ca
c_{2\beta}\big)+\lambda_7\cb\big(\sb\cab+\sa c_{2\beta}\big)\bigr]\,,\crr
&g\ls{H^0h^0h^0} = {-6\mw\over g_2}\bigl[
   \lambda_1\sa^2\cb\ca+\lambda_2\ca^2\sb\sa+\widetilde\lambda_3
   (\sa^3\sb+\ca^3\cb-\twothirds\cba)\cr
&\qquad\qquad\quad -\lambda_6\sa\big(\cb c_{2\alpha}+\ca\cab\big)
+\lambda_7\ca\big(\sb c_{2\alpha}+\sa\cab\big)\bigr]\,,\crr
&g\ls{H^0H^+H^-} =g\ls{H^0A^0A^0}-{2\mw\over g_2}
   \big(\lambda_5-\lambda_4\big)\cba\,,\crr
&g\ls{h^0H^+H^-} =g\ls{h^0A^0A^0}-{2\mw\over g_2}
   \big(\lambda_5-\lambda_4\big)\sba\,.\cr}
\eqn\defghaa
$$
(In our notation, $g_1\equiv g'$ and $g_2 \equiv g$.)
It is interesting to note that couplings of the charged Higgs bosons
satisfy relations analogous to that of $\mhpm$ given in eq.~\dchrgd.

\REF\pdg{K. Hisaka \etal\ [Particle Data Group], \sl Phys. Rev. \bf
D45\rm , S1 (1992).}%
\REF\otherlep{D. Decamp \etal\ [ALEPH Collaboration], {\sl Phys. Lett.}
{\bf B237}, 291 (1990);
P. Abreu \etal\ [DELPHI Collaboration], {\sl Phys. Lett.}
{\bf B245}, 276 (1990); \sl Nucl. Phys. \bf B373\rm , 3
(1992); B. Adeva \etal\
[L3 Collaboration], {\sl Phys. Lett.} {\bf B251}, 311 (1990);
M.Z. Akrawy \etal\ [OPAL Collaboration], {\sl Z. Phys.} {\bf C49},
1 (1991); P.D. Acton \etal, \sl Phys. Lett. \bf B268\rm , 122 (1991).}
\REF\alef{D. Decamp \etal\ [ALEPH Collaboration], {\sl Phys. Lett.}
{\bf B265}, 475 (1991).}%
At present, some experimental constraints on the
parameters of the two-Higgs doublet model have been obtained at LEP.
Here we briefly summarize the results for the Higgs search
as compiled by the Particle Data Group\refmark\pdg.
For the charged Higgs boson, $\mhpm>41.7$ GeV.
This is the most model independent bound and assumes only that
the $\hpm$ decays dominantly into $\tau^+\nu_\tau$, $c \bar s$
and $c\bar b$.
The LEP limits on the masses of $\hl$ and $\ha$ are obtained by searching
simultaneously for $Z \to h^0 f\bar f$ and $Z \to h^0 A^0$%
\refmark{\otherlep,\alef}. The $ZZ\hl$ and $Z\hl\ha$
couplings which govern these two decay rates are proportional to
$\sin(\beta-\alpha)$ and $\cos(\beta-\alpha)$, respectively.
Thus, one can use the
LEP data to deduce limits on $\mhl$ and $\mha$ as a function of
$\sin(\beta-\alpha)$\refmark\alef.
\REF\physlet{R. Hempfling, \sl Phys. Lett. \bf B296\rm , 121 (1992).}%
Stronger limits can be obtained in the MSSM where
$\sin(\beta-\alpha)$ is fixed by other model parameters.  The present
limits as summarized by the Particle Data Group\refmark\pdg\
are $\mhl>29$ GeV
and $\mha> 12$ GeV based on supersymmetric tree-level relations
among Higgs parameters, but with no assumption for the value of $\tanb$.
If leading log radiative corrections are incorporated and $\tanb>1$
is assumed, then recent results of the ALEPH Collaboration yield
$\mhl>41$ GeV and $\mha>20$ GeV (at 95\% CL). However, as was shown
in ref.~[\physlet] (and will be discussed briefly in section 7),
the limit on $\mhl$ may be substantially weaker if large
squark mixing is permitted.

The experimental information on the
parameter $\tanb$ is quite meager.
\REF\hewett{V. Barger, J.L. Hewett and R.J.N. Phillips, {\sl Phys.
Rev.} {\bf D41}, 3421 (1990).}
For definiteness, we shall assume that the Higgs-fermion couplings are
specified as in the MSSM.
In the Standard Model, the
Higgs coupling to top quarks
is proportional to $g_2m_t/2\mw$, and is
therefore the strongest of all Higgs-fermion couplings.
For $\tanb<1$, the Higgs couplings to top-quarks
in the two-Higgs-doublet model discussed above
are further enhanced by a factor of $1/\tanb$.  As a result, some
weak experimental limits on $\tanb$ exist based on the
non-observation of virtual effects involving the $H^-t\bar b$
coupling.  Clearly, such limits depend both on $\mhpm$ and $\tanb$.
For example, for $\mhpm\simeq\mw$, limits from the analysis of
$B^0$-$\overline{B^0}$ mixing
imply that $\tanb\gsim 0.5$\refmark\hewett.
No comparable limits exist based on top-quark couplings to neutral
Higgs bosons.

Theoretical constraints on $\tanb$ are also useful.  If $\tanb$
becomes too small, then the Higgs coupling to top quarks becomes
strong.  In this case, the tree-unitarity of processes involving the
Higgs-top quark Yukawa
coupling is violated.  Perhaps this should not be
regarded as a theoretical defect, although it does render any
perturbative analysis unreliable.  A rough lower bound advocated
by ref.~[\hewett], $\tanb\gsim m_t/600$~GeV, corresponds to a Higgs-top
quark coupling in the perturbative region.  A similar argument
involving the Higgs-bottom quark coupling would yield $\tanb\lsim 120$.
\REF\bagger{J. Bagger, S. Dimopoulos and  E. Masso, \sl Phys. Lett.
\bf 156B\rm , 357 (1985); \sl Phys. Rev. Lett. \bf 55\rm , 920 (1985).}
\REF\russ{G.M. Asatryan, A.N. Ioannisyan and S.G. Matinyan,
\sl Sov. J. Nucl. Phys. {\bf 53}, \rm 371 (1991) [\sl Yad. Fiz. {\bf 53},
\rm 592 (1991)];
M. Carena, T.E. Clark, C.E.M. Wagner, W.A. Bardeen and K. Sasaki,
{\sl Nucl. Phys.} {\bf D369}, 33 (1992);
H.E. Haber and F. Zwirner, unpublished.}
A more solid theoretical constraint is based on the
requirement that Higgs--fermion couplings remain finite when running
{}from the electroweak scale to
some large energy scale $\Lambda$\refmark{\bagger,\russ}.
Beyond $\Lambda$, one assumes that new physics enters.  The
limits on $\tanb$ depend on $\mt$ and
the choice of the high energy scale
$\Lambda$\refmark{\hiroshima,\bagger,\russ}.
\REF\cdflimit{F. Abe \etal\ [CDF Collaboration], {\sl Phys. Rev.
Lett.} {\bf 68}, 447 (1992).}%
For example, if there is no new physics (other than perhaps minimal
supersymmetry) below the grand unification scale of $10^{16}$~GeV,
then based on the CDF limit\refmark\cdflimit\ of $\mt>91$~GeV, one
would conclude that $0.5\lsim\tanb\lsim 50$.
\REF\tanbt{See \eg, G.F. Giudice and G. Ridolfi, \sl Z. Phys.
{\bf C41}, \rm 447 (1988); M. Olechowski and S. Pokorski, \sl
Phys. Lett. {\bf B214}, \rm 393 (1988); M. Drees and M.M. Nojiri,
{\sl Nucl. Phys.} {\bf B369}, 54 (1992).}
Finally, it is interesting to note that these limits on $\tanb$
are not very different from those that
emerge from models of low-energy supersymmetry based on supergravity
which strongly favor $\tanb>1$\refmark\tanbt.

\chapter{The Radiatively Corrected Higgs Sector of the MSSM}

In a general two-Higgs-doublet model none of the relations
derived in section 2 are
very predictive due to the large number of unknown parameters.
However, in
the MSSM, supersymmetry (SUSY) implies constraints among these
parameters, thereby leading to numerous predictions for Higgs masses and
coupling constants in terms of a few basic model parameters.

Consider first the case of unbroken SUSY.
Here all the Higgs self-coupling constants are related to the gauge
coupling constants
$$\eqalign{%
\lambda_1 &=\lambda_2 = \fourth (g_2^2+g_1^2)\,,\cr
\lambda_3 &=\fourth (g_2^2-g_1^2)\,,\cr
\lambda_4 &=-\half g_2^2\,,\cr
\lambda_5 &=\lambda_6=\lambda_7=0\,.\cr}
\eqn\bndfr$$
As a result the Higgs sector of the MSSM is
completely determined by two new measurable
quantities, which can be conveniently chosen to be $\mha$
and $\tanb$\refmark\hhg.

However, the parameters of
any theory will in general depend on the energy scale ($\sqrt{s}$) at
which they are evaluated. This dependence is described by the
renormalization group equations (RGEs)
$$ \d p_i/\d t
=\beta_i(p_1,p_2,..)\,,\qquad \hbox{ where}~t\equiv\ln(s)\,.
\eqn\rges$$
Here the parameters $p_i$ stand for the Yukawa couplings
$h_f$ $(f=u,d,\ell)$,
the gauge couplings of the $\smgaugegroup$~ gauge group $g_i$
($i=1,2,3$), the
Higgs self-coupling constants $\lambda_j$ ($j=1,...,7$),
and the mass parameters of the Higgs bosons $m_{ij}^2$ $(i,j=1,2)$.
Since eq.~\bndfr\ is valid at an arbitrary energy scale $\sqrt{s}$ we
find analogous relations for the corresponding $\beta$-functions
($\beta_{p_i}\equiv\beta_i$) by
taking the derivatives with respect to $t$
$$\eqalign{
&\beta_{\lambda_1}=\beta_{\lambda_2}= \fourth
[\beta_{g_2^2}+\beta_{g_1^2}]\,,\cr
&\beta_{\lambda_3}= \fourth
[\beta_{g_2^2}-\beta_{g_1^2}]\,,\cr
&\beta_{\lambda_4}=
-\half \beta_{g_2^2}\,.} \eqn\bndbeta$$
Note that eq.~\bndbeta\ is valid only if the theory is supersymmetric
at the scale $\sqrt{s}$.
However, if SUSY-breaking terms are
included, the mass degeneracy between the particles and their
supersymmetric partners
is violated, and the masses of the
superparticles can become heavy.
For simplicity we will typically assume that the supersymmetric particle
masses
are roughly of the same order.
That is, the scale of SUSY-breaking is characterized by one single
parameter,  $\msusy$. Then the $\beta$-functions at an intermediate
scale $\sqrt{s}$ (for $\mz<\sqrt{s}<\msusy$)
will no longer satisfy eq.~\bndbeta\ and
the gauge coupling constants and the self-coupling constants will evolve
differently. In the case of multiple SUSY-breaking scales,
as $\sqrt{s}$ decreases
one would have
to modify the $\beta$-functions
every time a
multiplet of supersymmetric particles decouples.
In this section we shall assume that the mass parameters of
the Higgs potential are of the order of $\mz$. This guarantees that both
Higgs doublets are present in the low effective theory
which is equivalent to the
(non-supersymmetric) SM with an extended two-doublet Higgs sector.
The required $\beta$-functions
are presented in Appendix A.
\foot{In refs.~[\barbi--\quiros],
it was assumed that $\mha\simeq\msusy$,
in which case the effective low-energy theory
consists of the non-supersymmetric SM with one physical Higgs boson.
Our approach extends the results of these authors by allowing
for the possibility
that all five physical Higgs states ($\hl, \hh, \ha$ and $\hpm$)
may have masses substantially below $\msusy$.}

\REF\hhh{A. Brignole and F. Zwirner,
{\sl Phys. Lett.} {\bf B299}, 72 (1993).}%
\REF\cpr{P.H. Chankowski, S. Pokorski and J. Rosiek,
MPI-Ph/92-116 (1992).}
\REF\hhhpierce{D. Pierce, to appear in the Proceedings of
the Santa Cruz workshop
on Electroweak Symmetry Breaking at Colliding-Beam Facilities, December,
1992.}
We begin by running the gauge coupling constants
{}from the electroweak scale (where they are measured)
up to $\msusy$. At this scale, the SUSY boundary conditions
given by eq.~\bndfr\ can be imposed, and we
obtain the values of the Higgs self coupling constants $\lambda_i$
at $\msusy$. Next we determine $\lambda_i(\mweak)$ by integrating the
corresponding RGEs from $\msusy$ down to $\mweak$.
Finally, using $\lambda_i$
in eqs.~\dchrgd--\defghaa, we find the RGE-improved Higgs masses and
trilinear interactions.\foot{In the leading logarithmic approximation,
the masses obtained in this manner are
physical masses (corresponding to the pole of the Higgs propagator).
Similarly, all one-loop definitions of the
three and four-point couplings differ only in their non-leading
logarithmic terms.}
RGE-improved Higgs masses based on the
two-Higgs doublet RGEs have also recently appeared in ref.~[\sasaki].
Analytic approximations to the trilinear Higgs interactions can
also be found in the literature.   In ref.~[\ellis], the $\hl\ha\ha$
coupling was computed using the effective potential method (in which
only terms explicitly proportional to $\mt^4$ were kept).  All other
three-Higgs interactions were obtained using the same approximation
scheme in ref.~[\bbsp].  In ref.~[\hhn], the $\hl\ha\ha$ coupling
was obtained by the method outlined above, in which the RGE-derived
expressions for $\lambda_i(\mweak)$ were inserted into eq.~\defghaa,
and is contrasted with the results of the effective potential
technique.  Since this work was completed, a number of more
complete computations based on one-loop vertex corrections to the
Higgs self-couplings have appeared.  The correction to the $hhh$
coupling can be found in refs.~[\hhh] and [\cpr];
see also ref.~[\hhhpierce].

By using the running parameters of the theory evaluated at
the electroweak scale ($\mweak$), one has
incorporated the leading logarithmic radiative corrections
to the Higgs parameters, summed to all orders in
perturbation theory. The RGEs can be solved by numerical
analysis using the computer. But it is instructive to solve
the RGEs iteratively. To first approximation we can take
the right hand side of eq.~\rges\ to be independent of
$\ln(s)$. That is, we evaluate the $\beta_i$ by
imposing tree-level relations among the parameters $p_i$
[\ie, eq.~\bndfr] and
evaluating the results at the scale
$\sqrt{s}=\mweak$. Then, integration of the RGEs is
trivial, and we obtain
$$
p_i(s_1)=p_i(s_2)-\beta_i\ln\left({s_2\over
s_1}\right)\,,\eqn\solvrge
$$
under the assumption that the particle content of the
effective low-energy theory does not change within the
range of integration.

The lower limit of integration is the electroweak
scale, $\mweak$. One possible choice for this scale is
$\mweak=\mz$ (which is the appropriate scale for
diagrams involving neutral gauge and Higgs bosons
inside the loops). A second equally reasonable choice
would be $\mweak=\mt$ (which is the appropriate scale
for diagrams involving the top-quark). This is a somewhat
arbitrary decision, since a different choice would yield results
that differ formally from eq.~\solvrge\ by a non-leading
logarithmic term (\ie, a term that does not grow as
$\ln(s_2)$ where $\sqrt{s_2}$ is the large scale). On the
basis of more precise one-loop calculations, we have adopted
the following strategy. We integrate the RGEs from
$\msusy$ down to $\mt$. At that point, we formally
integrate out the top-quark from the low-energy
theory, and finally integrate the appropriate RGEs of
the new low-energy effective theory down to $\mz$. The first
order solution of eq.~\rges\ becomes
$$
p_i(\mzz)=p_i(\msusyy)-\beta_i\ln\left({\msusyy\over\mtt}\right)
-\beta_i^0\ln\left({\mtt\over\mzz}\right)\,,\eqn\solvmzmt
$$
where the $\beta_i$ are the $\beta$-functions of the SM
with two Higgs doublets presented in Appendix A. The
$\beta_i^0$ are the $\beta$-functions in the same model
with the top-quark decoupled as we now explain.

We obtain $\beta_{\lambda_i}^0$ from $\beta_{\lambda_i}$
by setting the top-quark Yukawa coupling to zero.
For $\beta_{g_i^2}^0$, the situation is more subtle. The
reason is that below the top quark-threshold, the $\sutwouone$
gauge symmetry is broken. As a result, the coupling constants for
vertices involving the gauge bosons are no longer
constrained by the  $\sutwouone$
gauge symmetry and can evolve independently. Thus, to
determine the evolution of the $g_i$ we have to define
these couplings more precisely. Since our physical input
parameters are the masses of the $Z$ and $W$ bosons,
it is appropriate to define $g_1$ and $g_2$ through the
interaction
$$
\call=\fourth
\Big[\big(H_1^0\big)^2+\big(H_2^0\big)^2\Big]\Big(G^{\pm}
W_{\mu}^+W^{\mu-}+\half G^{ij}V_{\mu
i}V^{\mu}_j\Big)\eqn\defgg$$ where
$V_i=\big(W^3,B\big)$ are the neutral $\sutwol$ and $\uy$
gauge fields. The boundary conditions for the coupling
constants $G^{\pm}$ and $G^{ij}$ above the $\sutwouone$
breaking scale $\mt$ are
$$G^{\pm}=g_2^2,\qquad
G^{ij}=\bar g_i\bar g_j\qquad\hbox{with}~\bar
g_i=(g_1,-g_2)\,.\eqn\bndgg$$
After $\sutwouone$ symmetry
breaking, the set of $\beta$-functions becomes much larger
and more complicated. However, if we only work to first
order in perturbation theory below $\mt$,
\foot{That is, we do not solve the full set of RGEs below $\mt$.
For numerical purposes, it is certainly sufficient to isolate
the one-loop leading $\ln(\mtt/\mzz)$ terms.}
 it is not necessary to derive a full set of
$\beta$-functions. The only ones needed, $\beta_{G^\pm}$
and $\beta_{G^{ij}}$, are given in Appendix B. The gauge
boson masses are related to the coupling constants by
$$
\mww = \fourth v^2G^{\pm}(\mz)\,,\qquad\mzz = \fourth
v^2\tr G^{ij}(\mz)\,.\eqn\gmass
$$
One can solve the RGEs of Appendix B for $G^{ij}$ with the ansatz
$G^{ij}(t)=\bar g_i(t)\bar g_j(t)$
and therefore
$G^{ij}(t)$ is a rank one matrix for arbitrary $t$. Thus
the neutral gauge boson mass matrix maintains its zero mass
eigenvalue corresponding to the massless photon even below
the top quark-threshold.  Therefore, it is convenient to define
$g_1$ and $g_2$ in terms of $G^\pm(\mz)$ and $G^{ij}(\mw)$
$$\eqalign{
g_2^2 &\equiv G^\pm(\mz)\,,\cr
g_1^2 + g_2^2 &\equiv \tr\,G^{ij}(\mz)\,.\cr}\eqn\coupdef
$$

We now have to choose a value for $\msusy$. Of course the
simplest possibility is one in which all supersymmetric
particle masses are of order $\msusy$. For completeness we
briefly discuss the case of multiple mass scales in the
supersymmetric particle sector.
In the higgsino/gaugino sector we
have two free mass parameters $\mu$ and $M_2$ (it is common to
fix $M_1$ by the grand unification relation  $M_1 =
\fivethirds\tan^2\theta_W M_2$). We have computed the
contributions of the gauginos and the higgsinos to
$\beta_{\lambda_i}$ ($i=1,..,4$) and $\beta_{g_i^2}$ $(i=1,2$). It
is clear that the contributions of $\widetilde W$,
$\widetilde B$ and $\widetilde H$ to the $\beta_{g_i^2}$ can be
computed separately and have to be included at a scale
$\sqrt{s}>M_2$, $M_1$ and $\abs{\mu}$, respectively.\foot{We
choose a convention in which $M_2$ is positive.}
In the case of the
$\beta_{\lambda_i}$, both gauginos and higgsinos have to be
present at the scale $\sqrt{s}$ to yield a contribution. As a
result, the supersymmetric contributions to $\beta_{\lambda_i}$
that are proportional to $g_2^2$, $g_1^2$ or the product $g_1g_2$
have to be included if the scale
$\sqrt{s}>\mu_2\equiv\max\{\abs{\mu},M_2\}$,
$\sqrt{s}>\mu_1\equiv\max\{\abs{\mu},M_1\}$ or
$\sqrt{s}>\mu_{12}\equiv\max\{\mu_1,\mu_2\}$, respectively.
The gluino contributes only to $\beta_{g_3^2}$ at one-loop.
The gluino mass ($M_3$) is typically fixed by the grand
unification relation $M_3 = (g^2_3/g^2_2)M_2$.
Squark mass parameters are defined explicitly in section 6.
For simplicity
we neglect the possibility of generation mixing in the
squark mass matrices.
The result for $\beta_{\lambda_i}$
and $\beta_{g_i^2}$ for the MSSM are presented in Appendix A.

We can now compute the
effective low-energy coupling constants $\lambda_i(\mz)$
of the Higgs potential by evolving the coupling constants
in eq.~\bndfr\ from  $\msusy$ to $\mz$ by means of
eq.~\solvrge. Plugging those $\lambda_i$ into
eq.~\massmhh\ we find the elements of the CP-even Higgs
mass matrix. We
focus our attention on the simplest case where all supersymmetric
mass parameters are roughly degenerate. With the first
order leading log results for the $\lambda_i$ presented in
Appendix C these matrix elements become in this case
$$\eqalign{
\calm_{11}^2&=m_A^2\sb^2+m_Z^2\cb^2
+{g_2^2\mzz\cb^2\over96\pi^2\cw^2}\bigg[
P_t~\ln\left({\msusyy\over m_t^2}\right)\cr
&~~+\bigg(12N_c{m_b^4\over\mz^4\cb^4}-6N_c{m_b^2\over\mzz\cb^2}
+P_f+P_g+P_{2H} \bigg)\ln\left({\msusyy\over
m_Z^2}\right)\bigg] \cr \calm_{22}^2&=m_A^2\cb^2+m_Z^2\sb^2
+{g_2^2\mzz\sb^2\over96\pi^2\cw^2}\bigg[\bigg(P_f+P_g+P_{2H}
\bigg)\ln\left({\msusyy\over m_Z^2}\right)\cr
&~~+\bigg(12N_c{m_t^4\over\mz^4\sb^4}-6N_c{m_t^2\over\mzz\sb^2}
+P_t\bigg)\ln\left({\msusyy\over m_t^2}\right)\bigg]\cr
\calm_{12}^2&=-\sb\cb\bigg\{m_A^2+m_Z^2
+{g_2^2\mzz\over96\pi^2\cw^2}\bigg[\bigg(P_t-3N_c{m_t^2\over\mzz\sb^2}
\bigg)\ln\left({\msusyy\over m_t^2}\right)\cr
&~~+\bigg(-3N_c{m_b^2\over\mzz\cb^2}+P_f+P_g'+P_{2H}'
\bigg)\ln\left({\msusyy\over
m_Z^2}\right)\bigg]\bigg\}\cr}\eqn\mthree$$
where the constants $P_i$ are
$$\eqalign{P_t~&\equiv~~N_c(1-4e_u\sw^2+8e_u^2\sw^4)\,,\cr
P_f~&\equiv~~ N_g\big\{N_c[2-4\sw^2+8(e_d^2+e_u^2)\sw^4]
+[2-4\sw^2+8\sw^4]\big\}-P_t\,,\cr
P_g~&\equiv-44+106\sww-62\sw^4\,,\cr
P_g'~&\equiv~~10+34\sww-26\sw^4\,,\cr P_{2H}&\equiv
-10+2\sww-2\sw^4\,,\cr
P_{2H}'&\equiv~~8-22\sww+10\sw^4\,.\cr
}\eqn\defpp$$
Here
the subscripts $t, f, g$ and $2H$ correspond to the
contributions from the top-quark, the fermions (excluding the
top-quark), the gauge bosons and the two Higgs doublets.
By diagonalizing the CP-even mass matrix given
by eq.~\mthree\ one obtains the
individual neutral CP-even Higgs masses and
mixing angle $\alpha$ [eq.~\defalpha]. With the angle
$\alpha$ in hand, one then obtains the couplings of the
Higgs bosons to gauge bosons which are proportional to
either $\sinbma$ or $\cosbma$. Finally, the
$\lambda_i(\mz)$ and the angle $\alpha$ determine the
Higgs self couplings [eq.~\defghaa]. These results
were previously presented in ref.~[\hiroshima].
Similar results have also recently
been obtained in ref.~[\berkeley] using an effective
potential computation. However,
our results differ from those in ref.~[\berkeley] by terms of order
$g_2^2\mzz s_W^2\ln(\mtt/\mzz)$ due to our
more precise treatment of the theory
below the top-threshold [as explained Appendix B].

Of course the CP-even Higgs mass matrix may be computed
numerically by employing the $\lambda_i(\mz)$ obtained
through numerical solution of the RGEs. The resulting
Higgs mass matrix will then be the RGE-improved version
of eq.~\mthree, incorporating leading logarithmic effects
beyond one-loop order. In section 8 we will compare
these two results to see the numerical implications of
RGE-improvement.

\REF\gutu{J.F. Gunion and A. Turski, {\sl Phys. Rev.} {\bf D39}, 2701
(1989); {\bf D40}, 2333 (1989).}
\REF\brignole{A. Brignole, {\sl Phys. Lett.} {\bf B277}, 313 (1992).}
\REF\marco{M.A. D\'\i az and H.E. Haber, \sl Phys. Rev. \bf D45\rm ,
4246 (1992).}
Radiative corrections to the charged Higgs mass sum rule
have been obtained in refs.~[\gutu--\marco].
For completeness, we
also give the
one-loop leading logarithmic expression for the charged Higgs
mass. From eq.~\dchrgd\ and (C.6), we obtain
$$\eqalign{%
m_{H^{\pm}}^2&=m_A^2+m_W^2 +{{N_c g_2^2}\over{32\pi^2m_W^2}}
\Bigg[{{2m_t^2m_b^2}\over{s_{\beta}^2c_{\beta}^2}}-m_W^2
\bigg({{m_t^2}\over{s_{\beta}^2}}+{{m_b^2}\over{c_{\beta}^2}}\bigg)
+{\textstyle{2\over 3}}m_W^4\Bigg]
\ln\left({{\msusy^2}\over{m_t^2}}\right)  \cr
&\qquad+{g_2^2{m_W^2}\over{48\pi^2}}\left[N_c(N_g-1)+N_g+\half N_H-10
+15\tan^2\theta_W\right]
\ln\left({{\msusy^2}\over{m_W^2}}\right)\,,
\cr}\eqn\llform$$
where the number of Higgs doublets is $N_H = 2$.
The coefficient of $\ln(\msusy^2/\mt^2)$ in eq.~\llform\ is
consistent with the exact one-loop calculations of refs.~[\brignole]
and [\marco].  Note that this result differs slightly from the
one quoted in ref.~[\ellis], which is based on an effective
potential computation.  This is not surprising in light of the remarks
made in the Introduction.
As discussed in refs.~[\diaz] and [\brignole],
the effective potential is determined from Green functions evaluated
at zero external momenta, while physical masses are determined by
the pole of the (radiatively corrected) propagator.
This implies that for a particle
whose tree-level mass is nonzero, the effective potential does not
yield the complete leading logarithmic contribution to the particle
mass.

The one-loop leading log results [eqs.~\mthree--\llform]
are useful approximations (in the absence of large
squark mixing---see section 6) to the fully integrated
RGE results. This will be discussed more fully in section 8.
One clarification is necessary.
Consider the fully integrated RGE result for
some parameter $p_i$. Technically, we have only determined $p_i$ down to
$\sqrt{s}=\mt$. For $\sqrt{s}<\mt$,
we have only used the one-loop approximation [as in eq.~\solvmzmt]
to evolve $p_i(s)$ all the way down to $\sqrt{s}=\mweak$. Since $\mt$
cannot be very much larger than $\mz$, this procedure is certainly
sufficient for our purposes.
\endpage
\chapter{The Large $\bold\mha$ Limit}

Until now we have assumed that $\mha\simeq\calo(\mz)$.
In particular, the
low-energy effective theory (at the electroweak scale)
contains two Higgs doublets.  Consequently,
when we refer to the VEVs $v_i$ in section 3, we mean $v_i(\mz)$.
In this section we shall generalize our analysis to arbitrary
$\mha$. First we consider what happens when $\mha\gg\mz$. If we
expand eq.~\massev\ in powers of $\mzz/\mha^2$ we find
$$\eqalign{
\mhl^2&=\cb^2\calm_{11}^2+\sb^2\calm_{22}^2+2\sb\cb
\calm_{12}^2+\calo\left({\mz^4\over\mha^2}\right)\cr &=\mzz\ctwob^2
+{g_2^2m_Z^2\over96\pi^2\cw^2}
\Bigg\{\bigg[12N_c{m_b^4\over\mz^4}-6N_c\ctwob{m_b^2\over\mzz}
+\ctwob^2P_f\cr
&~~~+(P_{g}+P_{2H})(\sb^4+\cb^4)-2\sb^2\cb^2(P_{g}'+P_{2H}')
\bigg]\ln\left({\msusyy\over\mzz} \right)\cr
&\quad+\bigg[12N_c{m_t^4\over\mz^4}+
6N_c\ctwob{m_t^2\over\mzz}+\ctwob^2P_t\bigg]
\ln\left({\msusyy\over\mtt}\right)\Bigg\}\cr
&~~~+\calo\left({m^4_Z\over\mha^2}\right)
 +\calo\left[g_2^4\mzz\ln\left({\mha^2\over\mzz}\right)\right]
\,, }\eqn\mapprox$$
after using the results of eqs.~\mthree\ and \defpp. Since
eqs.~\mthree\ and \defpp\
were derived under the assumption
that $\mha\lsim\calo(\mz)$ it follows that the terms
proportional to $P_{2H}$ and $P_{2H}'$ will be modified when
$\mha\gg\mz$ by terms
of $\calo[g_2^2\mzz\ln(\mha^2/\mzz)]$ as indicated above.

Alternatively, we may investigate the case of large $\mha$
by integrating out the heavy Higgs doublet. In this
scenario one of the mass eigenvalues of $\calm_{ij}^2$ is much
larger than the weak scale. Then, in order to obtain the effective
Lagrangian at $\mweak$, we first have to run the various coupling
constants to the threshold $\mha$. Then we diagonalize the
Higgs mass matrix and express the Lagrangian in terms of the mass
eigenstates. Notice that in this case the mass eigenstate $\hl$ is
directly related to the field with the non-zero VEV  [\ie,
$\beta(\mha)=\alpha(\mha)+\pi/2+\calo(\mzz/\mha^2)$]. Below $\mha$
there remains only the SM Higgs doublet
$\phi\equiv\cb\Phi_1+\sb\Phi_2$. The potential is
$$
\calv=m_{\phi}^2(\phi^{\dagger}\phi)
+\half\lambda(\phi^{\dagger}\phi)^2
\,,\eqn\potsm
$$
where the boundary condition for $\lambda$ at $\mha$ is
$$
\eqalign{\lambda(\mha)&=\left[\cb^4\lambda_1+\sb^4\lambda_2+
2\sb^2\cb^2\widetilde\lambda_3
+4\cb^3\sb\lambda_6+4\cb\sb^3\lambda_7\right](\mha)\crr
&=\left[\fourth(g_1^2+g_2^2)\ctwob^2\right](\mha)
+{g_2^4\over384\pi^2\cw^4} \ln\left({\msusyy\over\mha^2}\right)\crr
&~~\times\bigg[12N_c\bigg({m_t^4\over\mz^4}+{m_b^4\over\mz^4}\bigg)
+6N_c\ctwob\bigg({m_t^2\over\mzz}-{m_b^2\over\mzz}\bigg)\cr
&\qquad+\ctwob^2\big(P_t+P_f\big)+(\sb^4+\cb^4)(P_g+P_{2H})
-2\sb^2\cb^2(P_g'+P_{2H}')\bigg]\,,
} \eqn\lapprox$$
where $(g_1^2+g_2^2)c_{2\beta}^2$is to be evaluated at the scale
$\mha$ as indicated. Similarly, the mass parameter $m_\phi$ that
appears in eq.~\potsm\ can be expressed in terms of the soft SUSY
mass parameters although we will not need this expression to compute
\REF\chengli{T.P. Cheng, E. Eichten and L.-F. Li, {\sl Phys. Rev.}
{\bf D9}, 2259 (1974).}
\REF\cmpp{N. Cabibbo, L. Maiani, G. Parisi, R. Petronzio
\sl Nucl. Phys. \bf B158, \rm 295 (1979).}
$\mhl$. The RGE in the SM for $\lambda$ is\refmark{\chengli,\cmpp}
$$
16\pi^2\beta_{\lambda} = 6\lambda^2
+\threeighth \left[2g_2^4+(g_2^2+g_1^2)^2\right]-2\sum_i
N_{c_i}h_{f_i}^4 -\lambda\left(\ninehalf
g_2^2+\threehalf g_1^2-2\sum_i N_{c_i}
h_{f_i}^2\right),
\eqn\defbetl $$
where the summation is over all fermions with
$h_{f_i}=gm_{f_i}/(\sqrt{2}\mw)$. The RGEs for the gauge couplings
are obtained from the $\beta_{g_i^2}$ given in Appendix A by putting
$N_H=1$. By solving the RGEs for scales $\mz<\sqrt{s}<\mha$
iteratively to first order we obtain
the light CP-even Higgs mass
$$
\eqalign{\mhl^2&=\lambda(\mz)v^2=\mzz\ctwob^2(\mha)\cr
&~~+{g_2^2m_Z^2\over96\pi^2\cw^2}
\Bigg\{\bigg[12N_c{m_t^4\over\mz^4}-
6N_c\ctwob^2{m_t^2\over\mzz}+\ctwob^2P_t\bigg]
\ln\left({\mha^2\over m_t^2}\right)\cr
&~~+\bigg[12N_c{m_b^4\over\mz^4}-6N_c\ctwob^2{m_b^2\over\mzz}
+\ctwob^2P_f+P_{1g}+P_{1H}\bigg]\ln\left({\mha^2\over\mzz}\right)\cr
&~~+\bigg[12N_c\bigg({m_t^4\over\mz^4}+{m_b^4\over\mz^4}\bigg)
+6N_c\ctwob\bigg({m_t^2\over\mzz}-{m_b^2\over\mzz}\bigg)
+\ctwob^2\big(P_t+P_f\big)\cr
&~~+(\sb^4+\cb^4)(P_g+P_{2H})-2\sb^2\cb^2(P_g'+P_{2H}')\bigg]
\ln\left({\msusyy\over\mha^2}\right)\Bigg\}
+\calo\left({m^4_Z\over\mha^2}\right).\cr
}\eqn\mhltt$$
Here we have defined
$$\eqalign{
P_{1H}\equiv&-9\ctwob^4+\left(1-2\sww+2\sw^4\right)\ctwob^2\,,\crr
P_{1g}\equiv&
\ctwob^2(-17+70\sww-44\sw^4)-(27-36\sww+18\sw^4)\cr &=
(\cb^4+\sb^4)P_{g}-2\sb^2\cb^2P_{g}'\,.\cr }\eqn\defpps$$
where
the subscripts $1H$ and $1g$ indicate that these are the Higgs and
gauge boson contributions in the one-Higgs-doublet model. However,
we still must deal with implicit scale dependence of $c_{2\beta}^2$.
Since the fields $\Phi_i$ $(i=1,2)$ change with the
scale\refmark{\brig,\drees}, it
follows that $\tanb$ scales like the ratio of the two Higgs doublet
fields, \ie,
$$
{1\over\tan^2\beta}{\d\tan^2\beta\over\d
t}={\Phi_1^2\over\Phi_2^2}{\d\over\d t}
\left({\Phi_2^2\over\Phi_1^2}\right)
=\gamma_2-\gamma_1\,.\eqn\deftanb $$
Thus we arrive at the RGE for $\cos 2\beta$ in terms of the
anomalous dimensions $\gamma_i$ given in eq.~(A.10)
$$
\ctwob^2(\mha)=\ctwob^2(\mz)
+4\ctwob\cb^2\sb^2(\gamma_1-\gamma_2)
\ln\left({\mha^2\over\mzz}\right)\,.\eqn\rgecb $$
Inserting eq.~\rgecb\ in eq.~\mhltt, we end up with
$$
\eqalign{
\mhl^2&= \mzz\ctwob^2(\mz)
+{g_2^2m_Z^2\over96\pi^2\cw^2}
\Bigg\{\bigg[12N_c{m_b^4\over\mz^4}-6N_c\ctwob{m_b^2\over\mzz}
+\ctwob^2P_f\cr
&~~+\left(P_{g}+P_{2H})(\sb^4+\cb^4\right)
 -2\sb^2\cb^2\left(P_{g}'+P_{2H}'\right)
 \bigg]\ln\left({\msusyy\over\mzz} \right)\cr
&~~+\bigg[12N_c{m_t^4\over\mz^4}+
 6N_c\ctwob{m_t^2\over\mzz}+\ctwob^2P_t\bigg]
 \ln\left({\msusyy\over m_t ^2}\right)\cr
&~~-\bigg[\left(\cb^4+\sb^4\right)P_{2H}-2\cb^2\sb^2P_{2H}'-P_{1H}\bigg]
 \ln\left({\mha^2\over\mzz}\right)\Bigg\}
 +\calo\left({m^4_Z\over\mha^2}\right)\,,
}\eqn\mhltot$$
which agrees with eq.~\mapprox. In particular, we have now obtained
explicitly the term proportional to $\ln(\mha^2)$ which
accounts for the fact that there are two Higgs doublets present at
a scale above $\mha$ but only one Higgs doublet below $\mha$.
Finally, by taking the limit of large $\tanb$, we have checked that
eq.~\mhltot\ reduces to the one-loop leading log result of
ref.~[\leter].

\chapter{Physical Definition of {$\bold\tanb$}}

In eq.~\polar\ we implicitly defined $\tanb$ in terms of VEVs evaluated
at the electroweak scale. However, this definition is not appropriate
in the case where $\mha\gg\mweak$, since the
definition of $\tanb$ requires the presence of both Higgs doublets
in our low-energy effective theory. Consider for example the
tree-level relation for the partial hadronic width of $A^0$
$$
\Gamma(\ha\rarrow b\bar b)={N_cg_2^2\mbb\over32\pi\mww}
\big(\mha^2-4\mbb\big)^{1/2}\tan^2\beta\,.\eqn\defgabb
$$
This tree-level relation is true for an arbitrary two-Higgs-doublet
model under the assumption that down-type (up-type) fermions
couple exclusively to $H_1$ ($H_2$). Using running parameters
evaluated at $\mha$, eq.~\defgabb\ continues to hold
even after
leading log corrections are
included.\foot{ However, there are non-leading log
corrections that generate Yukawa couplings of up-type (down-type)
fermions to $H_{1}$ ($H_2$).}
{}From eq.~\defgabb\
we can obtain a practical definition of
$\tanb(\mha)$. By using the RGE for $\tanb$ given in
eq.~\deftanb\ we can obtain $\tanb(\mz)$ which at the leading log
level matches the definition of $\tanb$ in terms of VEVs given in
eq.~\polar.

It is instructive to show that our results do not depend on the
physical definition of $\tanb$.  (See ref.~[\brig] for a detailed
discussion of the relation between different $\tanb$ definitions.)
Consider an alternative definition of
$\tanb$ advocated in ref.~[\yam] based on
the supersymmetric
tree-level relation
$$
\Delta m_{\widetilde
L}^2\equiv m_{\widetilde e}^2-m_{\widetilde\nu}^2=\mww\cos
2\beta+\calo\left({m_e^2\over\mzz}\right)\,.\eqn\defbeta $$
We can easily obtain the radiative corrections due to the
quark/squark sector, for arbitrary $\mha$. The relevant terms of the
Lagrangian are
$$\eqalign{\calv_{\widetilde L}&=\left[M_{\widetilde L}^2
+\Lambda_{ij}^L\left(\Phi^{\dag}_i\Phi_j\right)\right]
\widetilde L^{\dag}\widetilde L+\bar\Lambda_{ij}^L
\left(\Phi_i^{\dag}\widetilde L\right)\left(\widetilde
L^{\dag}\Phi_j\right)
\,,}\eqn\potslep
$$
where $\widetilde L\equiv(\widetilde\nu,\widetilde e_L)$ and the
indices $i,j=1,2$ run over the two Higgs doublet fields. In the MSSM
the tree-level quartic coupling constants $\Lambda$ and
$\bar\Lambda$ can be expressed in terms of the gauge
coupling constants. Clearly, only
the terms in eq.~\potslep\ proportional to $\bar\Lambda_{ij}^L$ cause a
selectron-sneutrino mass splitting. The supersymmetric boundary
conditions for $\bar\Lambda_{ij}^L$ are
$$
\bar\Lambda^L_{ij}=\half g_2^2\,\diag\left(-1,1\right)\,.\eqn\bndlep
$$

First, we note that there are no vertex corrections due to the
fermions (since we can neglect all Yukawa couplings except $h_b$
and $h_t$). Thus the contributions of the fermions to the
$\beta$-functions are
$$
\beta_{\bar\Lambda_{ij}^L}^f=-\bar\Lambda_{ij}^L\gamma_i^f\,,
\qquad\hbox{where}\qquad\cases{16\pi^2\gamma_1^f =- N_c
h_b^2\,,\cr 16\pi^2\gamma_2^f =- N_c h_t^2\,.}\eqn\betaqt
$$
Note that $\bar\Lambda_{ij}^L$ will remain diagonal at scales below
$\msusy$.
After removing the heavy Higgs doublet via
$$
\Phi_1\to\cb\phi\,,\qquad\Phi_2\to\sb\phi\,,\eqn\defphism
$$
we obtain $\Delta m_{\widetilde L}^2 = \half v^2 \bar\Lambda$, where
$\bar\Lambda$ is the coefficient of $|L^{\dag}\phi|^2$ in the
effective scalar potential at a scale $\sqrt{s}<\mha$.
The
$\beta$-function and the boundary condition for $\bar\Lambda$ at the
scale $\mha$ are $$ \eqalign{&
 \bar\Lambda(\mha) =
\left[\cb^2\bar\Lambda_{11}^L+\sb^2\bar\Lambda_{22}^L\right](\mha)\,,\cr
 &\beta_{\bar\Lambda} =
-\bar\Lambda\gamma^f =
-\bar\Lambda\left(\cb^2\gamma_1^f + \sb^2\gamma_2^f\right)\,.
}\eqn\blabla$$
Thus the resulting slepton squared mass difference is obtained
immediately
$$\eqalign{
\Delta m_{\widetilde L}^2 &=
\half v^2\bar\Lambda(\mweak)
=\mww\left[c_{2\beta}(\mha)+{\beta_{g_2^2}\over g_2^2}c_{2\beta}
\ln\left({\msusyy\over\mweakk}\right)\right.\cr
&\left.-\left(\cb^2\gamma_1-\sb^2\gamma_2\right)
\ln\left({\msusyy\over\mha^2}\right)
-c_{2\beta}\left(\cb^2\gamma_1+\sb^2\gamma_2\right)
\ln\left({\mha^2\over\mweakk}\right)\right]\,.}\eqn\dmhh
$$
This equation, which represents the one-loop leading log radiative
corrections to the slepton mass difference due to the quark/squark
sector, is sufficient for our purposes.
Note that the physical quantities in eq.~\dmhh\ [\ie,
$\mw$ and $\Delta m_{\widetilde L}^2$] cannot depend on the
arbitrary scale $\mha$.
It is a simple exercise to check that by taking the derivative of
eq.~\dmhh\ with respect to $\ln(\mha^2)$ we recover the
RGE for $\cos 2\beta$ [eq.~\deftanb].

\chapter{Radiative Corrections due to Soft Squark Interactions}

In the last two sections we have studied the leading log
corrections to the Higgs self coupling constants $\lambda_i$.
By inspecting the Lagrangian we find
that all dimension-four operators of the MSSM respect
two $\uone$ symmetries. Under these symmetries, the Higgs fields transform
as $\Phi_i\to e^{i\alpha_i}\Phi_i$, $i=1,2$. One combination of these
symmetries  is gauged [namely $\uy$] whereas the other one is a
global symmetry [which we call $\ug$] which imposes
constraints on the parameters of the theory. One result of the
$\ug$ symmetry is that the $\lambda_i$
($i = 5, 6, 7$) remain zero even after one-loop leading log
corrections are included. However, the $\ug$ symmetry is broken
by dimension-two and dimension-three terms of the MSSM.
The dominant corrections
derive from the squark mixing effects in the top and bottom sector.
These effects would lead for example to finite (non-logarithmic)
renormalizations of $\lambda_5, \lambda_6$ and $\lambda_7$.
In this section, we show how to obtain
such corrections (see also ref.~[\yama]).

\REF\susyhaber{For further details on the parameters of the MSSM,
see H.E. Haber, SCIPP-92/33 (1993),
to appear in the Proceedings of the 1992
Theoretical Advanced Study Institute in Elementary Particle Physics.}
The most general scalar potential (including Higgs fields and one
generation of squarks\foot{Contributions from the
sleptons and the other two squark generations are omitted in the
formulae presented in this section.})
takes the following form:
$$
\calv^0 = \calv_M + \calv_\Gamma + \calv_\Lambda +
\calv_{\widetilde Q}\,,\eqn\blabla
$$
where we have defined
$$\eqalign{
&\calv_M = (-1)^{i+j}m_{ij}^2\Phi_i^{\dag}\Phi_j+
M_{\widetilde Q}^2\left(\widetilde
  Q^{\dag}\widetilde Q\right)
+M_{\widetilde U}^2\widetilde U^*\widetilde U
+M_{\widetilde D}^2\widetilde D^*\widetilde D
\,,\cr
&\calv_\Gamma =
\Gamma_i^D\left(\Phi^{\dag}_i\widetilde Q\right)\widetilde D
+\Gamma_i^U\left(i\Phi_i^T\sigma_2\widetilde Q\right)\widetilde U\cr
&\calv_\Lambda =
\Lambda_{ik}^{jl}\left(\Phi^{\dag}_i\Phi_j\right)
\left(\Phi^{\dag}_k\Phi_l\right)
+\left(\Phi^{\dag}_i\Phi_j\right)
\left[\Lambda_{ij}^Q\left(\widetilde Q^{\dag}\widetilde Q\right)
+\Lambda_{ij}^U \widetilde U^*\widetilde U
+\Lambda_{ij}^D \widetilde D^*\widetilde D\right]\cr
&\qquad+\bar\Lambda_{ij}^Q\left(\Phi^{\dag}_i\widetilde Q\right)
\left(\widetilde Q^{\dag}\Phi_j\right)
+\half\left[\Lambda\epsilon_{ij}
\left(i\Phi_i^T\sigma_2\Phi_j\right)\widetilde
D^*\widetilde U+\hc\right]
\,,}\eqn\potsqu$$
and $\calv_{\widetilde Q}$ contains quartic squark interaction terms.
Henceforth, we discard $\calv_{\widetilde Q}$ since the
contributions of these terms to the finite renormalizations of the
$\lambda_i$ enter only at two-loop order.   In eq.~\potsqu,
$i,j,k,l = 1,2$ run over the two Higgs
doublet fields, the third generation squark fields are defined below
eq.~(A.8), and the various $\Gamma$'s and $\Lambda$'s are
determined by the tree-level SUSY relations
$$\eqalign{
\Lambda^Q&= \diag\left[\fourth(g_2^2-g_1^2Y_Q),
h_U^2-\fourth(g_2^2-g_1^2Y_Q)\right]\,,\cr
\bar\Lambda^Q&= \diag\left(h_D^2-\half
g_2^2, \half g_2^2-h_U^2\right)\,,\cr
\Lambda^U&= \diag\left(-\fourth g_1^2Y_U,
h_U^2+\fourth g_1^2Y_U\right)\,,\cr
\Lambda^D&= \diag\left(h_D^2-\fourth
g_1^2Y_D, \fourth g_1^2Y_D\right)\,,\cr
\Lambda~&=-h_Uh_D\,,}\eqn\blabla
$$
where $h_U$, $h_D$, and the squark hypercharges are given in
eq.~(A.9) and subsequent text, and
$$\eqalign{&\Gamma^U=h_U\left(-\mu, A_U\right)\,,\cr
&\Gamma^D=h_D\left(A_D, -\mu\right)\,,}\eqn\defgammam
$$
which define the $A$-parameters: $A_U$ and $A_D$.  (The parameter
$\mu$ also appears in the chargino/neutralino sector\refmark\susyhaber.)
The $\Lambda_{ik}^{jl}$ can be expressed easily in terms of the
$\lambda_i$ ($i= 1, 2,...$) of eq.~\pot.
If we introduce the collection of fields
$\Psi\equiv(\widetilde Q, \widetilde U^*, \widetilde D^*)$ and
use the fact that
$$
{\partial^2\calv^0\over\partial\Psi_a\partial\Psi_b} =
{\partial^2\calv^0\over\partial\Psi^*_a\partial\Psi_b^*} =
0\,\eqn\blabla $$
we obtain the squark mass matrix
$$
\calm^2_{ab}
={\partial^2\calv^0\over\partial\Psi_a\partial\Psi_b^*}\,.\eqn\blabla
$$

\REF\dimred{W. Siegel, {\sl Phys. Lett.} {\bf 84B}, 193 (1979);
{\bf 94B}, 37 (1980); D.M. Capper, D.R.T. Jones and
P. van Nieuwenhuizen, {\sl Nucl. Phys.} {\bf B167} (1980) 479.}
The RGEs due to the
squark sector presented in Appendix A can now be computed by
demanding that the one-loop renormalized potential
(in the Landau gauge using dimensional reduction\refmark\dimred\ and
the $\overline{MS}$-scheme)
$$
\calv = \calv^0 +
{N_c\over32\pi^2}\tr\calm^4
\left[\ln\left({\calm^2\over\sigma^2}\right)-\threehalf\right]
\eqn\poteff$$
is independent of the arbitrary renormalization scale
$\sigma$. Here, $N_c=3$ colors, and we have included a factor of 2 since
the squark fields are complex.  The
RGEs for the quartic Higgs couplings are
$$\eqalign{32\pi^2{\d\Lambda_{ik}^{jl}\over\d t}
&= N_c\left[
2\Lambda_{ij}^Q\Lambda_{lk}^Q+\bar\Lambda_{ij}^Q\Lambda_{lk}^Q
+\Lambda_{ij}^Q\bar\Lambda_{lk}^Q
+\bar\Lambda_{il}^Q\bar\Lambda_{jk}^Q\right.\crr
&\qquad +\left.\Lambda_{ij}^U\Lambda_{lk}^U+\Lambda_{ij}^D\Lambda_{lk}^D
+\Lambda^2\left(\delta_{ij}\delta_{lk}-\delta_{il}\delta_{jk}\right)
\right]\,.}\eqn\betagamma$$

The $\beta$-functions at scales below the mass
of one or more squark fields are obtained from eq.~\betagamma\ by
removing the contributions corresponding to these fields and assuming
that the coupling constants are continuous. It is
clear that the dimensionful coupling constants in eq.~\defgammam\
cannot contribute to the $\beta$-functions of dimensionless couplings
at scales larger than all the mass parameters in
$\calv^0$.
However, in the
presence of dimension-three terms the decoupling of heavy squarks
becomes non-trivial.
To understand what is happening, consider the path integral
derivation of eq.~\poteff, in the case of $M_{\widetilde
D}\gg M_{\widetilde U},M_{\widetilde Q}$ (\ie, we integrate out
$\widetilde D$; other cases are completely analogous). The generating
functional is
$$
W\propto\int\left[\d\widetilde D^*\d\widetilde
D\right]\exp\left\{i\int\d^4x\left[\widetilde
D^*(-i\Delta)^{-1}\widetilde D+\Gamma_i^D(\Phi_i^*\widetilde
Q)\widetilde D+\hc+\call_\Phi\right]\right\}\,,\eqn\pathint
$$
where all other terms of the Lagrangian are included in
$\call_{\Phi}$. The inverse propagator in the presence of
non-zero Higgs fields is $$
(-i\Delta)^{-1}=\dalam-\left[M_{\widetilde D}^2+h_D^2\Phi_1^2
+\fourth Y_Qg_1^2(\Phi_1^2-\Phi_2^2)\right]\,.\eqn\blabla
$$
The path integral over $\widetilde D$ and $ \widetilde D^*$ is
straightforward with the result
$$
W\propto\exp\left\{-i\int\d^4x
\Gamma_i^D(\Phi_i^*\widetilde Q)(-i\Delta)\left[
\Gamma_j^D(\Phi_j^*\widetilde Q)\right]^*+\call_\Phi\right\}
\,.\eqn\pathresult
$$
When external momenta are much smaller than $M_{\widetilde D}$,
the interaction term in eq.~\pathresult\
becomes local and can be absorbed into the scalar potential of the
low-energy effective theory by redefining
$$
\Lambda_{ij}^Q\to\Lambda_{ij}^{Q\prime}\equiv\Lambda_{ij}^Q-{1\over
M_{\widetilde D}^2}\Gamma_i^D\Gamma_j^D\,.\eqn\blabla
$$
These are inserted into the $\beta$-function
[eq.~\betagamma] which is used
to run the Higgs self-couplings at scales below $M_{\widetilde
D}$. The logs arising from such a calculation are only of
$\calo[\ln(M_{\widetilde U}^2/M_{\widetilde D}^2)]$. However, notice that
$\Lambda_{ij}^Q$ is
no longer diagonal and as a result the $\lambda_i$ $(i= 5, 6, 7)$ can
become non-zero. This can generate phenomena that are absent at the leading
log level and may be phenomenologically important in some circumstances.

The dimension-three terms also lead to corrections which have no
logarithmic dependence on the mass parameters of the model.
Consider the case of $M_{\widetilde U}=M_{\widetilde D}=M_{\widetilde Q}
\equiv\msusy$. In this case the
corrections described above are not present.
Nevertheless, important non-logarithmic corrections to the
$\Lambda_{ik}^{jl}$ can arise which lead to finite shifts in the
$\lambda_i$ (denoted by $\Delta\lambda_i$ below). These can be computed
by expanding the effective potential [eq.~\poteff] to fourth order in
$\Phi$ as described in Appendix D. There are two types of corrections
corresponding to triangle diagrams and box diagrams. The results for the
triangle diagrams (with two powers of trilinear coupling constants
$A\ls{U}$, $A\ls{D}$ or $\mu$) denoted by a superscript $(3)$ are
{\def\crr{\cr\noalign{\vskip 10pt}}
$$\eqalign{
&\Delta\lambda_1^{(3)} = {N_c\over16\pi^2\msusyy}\biggl\{
A\ls{D}^2h\ls{D}^2\bigg(2h\ls{D}^2-{g_2^2+g_1^2\over4}\bigg)+\mu^2h\ls{U}^2
{g_2^2+g_1^2\over4}\biggl\}\crr &
\Delta\lambda_2^{(3)} =
{N_c\over16\pi^2\msusyy}\biggl\{
A\ls{U}^2h\ls{U}^2\bigg(2h\ls{U}^2-{g_2^2+g_1^2\over4}\bigg)+\mu^2h\ls{D}^2
{g_2^2+g_1^2\over4}\biggl\}\crr
&
\Delta\lambda_3^{(3)} =
{N_c\over32\pi^2\msusyy}\biggl\{\mu^2\left(h\ls{U}^2-h\ls{D}^2\right)^2
+h\ls{U}^2h\ls{D}^2\left(A\ls{U}+A\ls{D}\right)^2\cr
&
\qquad\qquad\qquad+{g_1^2-g_2^2\over4}\left[\left(A\ls{D}^2-\mu^2\right)h\ls{D}^
2
+\left(A\ls{U}^2-\mu^2\right)h\ls{U}^2\right]\biggl\}\crr
&\Delta\lambda_4^{(3)} = {N_c\over32\pi^2 \msusyy}
\biggl\{\mu^2\left(h\ls{U}^2+h\ls{D}^2\right)^2
-h\ls{U}^2h\ls{D}^2\left(A\ls{U}+A\ls{D}\right)^2\cr
&
\qquad\qquad\qquad+{g_2^2\over2}\left[\left(A\ls{D}^2-\mu^2\right)h\ls{D}^2
+\left(A\ls{U}^2-\mu^2\right)h\ls{U}^2\right]
 \biggl\}\crr
&
\Delta\lambda_5^{(3)} = 0\cr
&
\Delta\lambda_6^{(3)} =
{N_c\mu\over32\pi^2 \msusyy}\biggl\{
A\ls{D}h\ls{D}^2\bigg({g_2^2+g_1^2\over4}-2h\ls{D}^2\bigg)
-A\ls{U}h\ls{U}^2{g_2^2+g_1^2\over4}
\biggl\}\crr
&
\Delta\lambda_7^{(3)} = {N_c\mu\over32\pi^2 \msusyy}\biggl\{
A\ls{U}h\ls{U}^2\bigg({g_2^2+g_1^2\over4}-2h\ls{U}^2\bigg)
-A\ls{D}h\ls{D}^2{g_2^2+g_1^2\over4}
\biggl\}}\eqn\dlmbdv$$
and the results
for the box diagrams (with four powers of $A\ls{U}$, $A\ls{D}$ or
$\mu$) denoted with a superscript (4) are
\endpage
$$\eqalign{
&\Delta\lambda_1^{(4)} =-{N_c\over96\pi^2\msusy^4}\biggl\{
A\ls{D}^4h\ls{D}^4+\mu^4h\ls{U}^4\biggl\}\crr
&\Delta\lambda_2^{(4)}
=-{N_c\over96\pi^2\msusy^4}\biggl\{
A\ls{U}^4h\ls{U}^4+\mu^4h\ls{D}^4\biggl\}\crr
&\Delta\lambda_3^{(4)}
 =-{N_c\over96\pi^2\msusy^4}\biggl\{
 \mu^2A\ls{U}^2h\ls{U}^4 + \mu^2A\ls{D}^2h\ls{D}^4 + h_U^2 h_D^2
 (\mu^2-A_UA_D)^2 \biggl\}\crr
&\Delta\lambda_4^{(4)} =-{N_c\over96\pi^2\msusy^4}\biggl\{
 \mu^2A\ls{U}^2h\ls{U}^4 + \mu^2A\ls{D}^2h\ls{D}^4 - h_U^2 h_D^2
 (\mu^2-A_UA_D)^2 \biggl\}\crr
&\Delta\lambda_5^{(4)} =-{N_c\mu^2\over96\pi^2\msusy^4}\biggl\{
A\ls{D}^2h\ls{D}^4+A\ls{U}^2h\ls{U}^4\biggl\}\crr
&\Delta\lambda_6^{(4)}
=\phantom{+} {N_c\mu\over96\pi^2\msusy^4}\biggl\{
\mu^2A\ls{U}h\ls{U}^4+A\ls{D}^3h\ls{D}^4\biggl\}\crr
&\Delta\lambda_7^{(4)}
=\phantom{+} {N_c\mu\over96\pi^2\msusy^4}\biggl\{
\mu^2A\ls{D}h\ls{D}^4+A\ls{U}^3h\ls{U}^4\biggl\}\,.
}\eqn\dlmbdav$$
}

Finally, the self energy diagrams yield corrections to the
kinetic term of the Higgs fields which have to be absorbed by
redefining the Higgs fields
$$
\Phi_i\to\widehat\Phi_i\equiv\left(\delta_{ij}-\half {\rm
A}_{ij}'\right)\Phi_j\,.\eqn\redefphi
$$
For example, the contributions to the $A_{ij}^\prime$ coming from the
trilinear scalar interactions are given by
$$
A_{ij}^\prime = -{N_c\over
96\pi^2\msusyy}\Bigg[h\ls{U}^2\left(\matrix{\mu^2 &-\mu
A\ls{U}\cr-\mu A\ls{U} &A\ls{U}^2\cr}\right)
+h\ls{D}^2\left(\matrix{A\ls{D}^2
&-\mu A\ls{D}\cr -\mu A\ls{D} &\mu^2\cr}\right)\Bigg]\,.\eqn\waver
$$
If we then express the quartic terms of the
potential in terms of the new fields $\widehat\Phi$ we obtain
\endpage
$$\eqalign{
&\Delta\lambda_1^{(2)} = \half(g_1^2+g_2^2)A_{11}'\,,\cr
&\Delta\lambda_2^{(2)} = \half(g_1^2+g_2^2)A_{22}'\,,\cr
&\Delta\lambda_3^{(2)} =-\fourth(g_1^2-g_2^2)
\big(A_{11}'+A_{22}'\big)\,,\cr
&\Delta\lambda_4^{(2)} =-\half g_2^2\big(A_{11}'+A_{22}'\big)\,,\cr
&\Delta\lambda_5^{(2)} = 0\,,\cr
&\Delta\lambda_6^{(2)} = \eighth(g_1^2+g_2^2)A_{12}'\,,\cr
&\Delta\lambda_7^{(2)} = \eighth(g_1^2+g_2^2)A_{12}'
\,,\cr}\eqn\dlbdaw
$$
where we have used the (supersymmetric)
tree-level results for the  $\lambda_i$ that
appeared on the right hand side of eqs.~\dlbdaw\ evaluated at the
scale $\msusy$.

Combining all the results obtained in this section, we conclude that
the appropriate boundary conditions for the $\lambda_i$ at $\msusy$ are
obtained by setting
$$
\lambda_i(\msusy) = \lambda_i({\rm SUSY})+\Delta\lambda_i^{(2)}
+\Delta\lambda_i^{(3)}+\Delta\lambda_i^{(4)}\,,\eqn\bndtot
$$
where $\lambda_i({\rm SUSY})$ are the scalar self-couplings
of the unbroken supersymmetric theory given in eq.~\bndfr\
evaluated at $\msusy$.
The low-energy effective Higgs potential is given by $\calv$ in
eq.~\pot\ where the $\lambda_i$ appearing there are $\lambda_i(\mweak)$
obtained by solving the RGEs [given in Appendix A] subject to the
boundary condition given in eq.~\bndtot.
Using the coupling constants so obtained, one may compute the mass
eigenvalues, mixing angles and coupling constants in
eq.~\dchrgd--\defghaa\ numerically.   The effective potential
formalism (see, \eg, refs.~[\rest,\ellis,\berkeley, and \drees])
correctly reproduces the terms that arise from $\Delta\lambda_i^{(3)}$
and $\Delta\lambda_i^{(4)}$.  However, the effective potential
method does not pick up the
terms arising from $\Delta\lambda_i^{(2)}$, which derive
{}from wave function renormalization.

In the next section we shall present numerical results of this procedure.
However, it
is instructive to first look at a few
analytic results. In the one-loop logarithmic approximation, the RGEs are
solved
analytically as shown in section 3. The effect of the new boundary conditions
[eq.~\bndtot] at the one-loop level is simply additive.  We shall include
only the effects of the third generation squark mixing.  If we
denote the one-loop leading log squared mass shifts obtained from
eq.~\mthree\ by $(\Delta m^2)_{\rm 1LL}$,
then we obtain the following
expressions for the neutral Higgs masses
in the limit of large $\tanb$:
$$
\eqalign{&\left(\mhl^2-\mzz\right)_{\beta=\pi/2}
= (\Delta\mhl^2)_{\rm 1LL} -{N_cg_2^2\mww\over
96\pi^2\msusy^4}\left[
\left({A_t\mt\over\sb\mw}\right)^4+\left({\mu\mb\over\cb\mw}\right)^4\right]\cr
&\qquad\qquad+{N_cg_2^2\mww\over96\pi^2\msusyy}\left[
12{A_t^2\mt^4\over\sb^4\mw^4}-4{A_t^2\mtt\over\sb^2\mww\cww}
+2{\mu^2\mbb\over\cb^2\mww\cww}\right]\,,}\eqn\dmanl
$$
and
$$
\left(\mhh^2-\mha^2\right)_{\beta=\pi/2}=
-{N_cg_2^2\mu^2\over96\pi^2\msusy^4}
\left({A_t^2\mt^4\over\sb^4\mww}
+{A_b^2\mb^4\over\cb^4\mww}\right)\,,\eqn\dmanh
$$
assuming that $\mha>\mz$ [if $\mha<\mz$ then interchange $\mhl$ and
$\mhh$]. Taking the limit $\beta\to\pi/2$ on the right hand side of
eqs.~\dmanl\ and \dmanh\ is subtle because of factors of $\cb$ in the
denominator. The appropriate limit is one where the Yukawa coupling
$h_b\equiv g_2\mb/(\sqrt2\mw\cb)$ is fixed. Thus, in the limit
$\beta\to\pi/2$, it follows that $\mb\to0$ such that $\mb/\cb$ is
fixed. We also refrain from setting $\sb=1$ to preserve the
symmetry of the formulae above.

As a second example consider the radiative corrections
to charged Higgs mass.  Define
$$
\Delta\mhpm^2\equiv \mhpm^2-\mha^2-\mww\,,
\eqn\defdpm$$
so that
$\Delta\mhpm^2 = 0$ at tree-level. We then find $$
\eqalign{\Delta\mhpm^2 &= (\Delta\mhpm^2)_{\rm 1LL}
 +{N_cg_2^2\mww\over96\pi^2}
  \left[{\mtt\over\mww\sb^2}\left({\mu^2-2A_t^2\over\msusyy}\right)
  +{\mbb\over\mww\cb^2}\left({\mu^2-2A_b^2\over\msusyy}\right)\right]\crr
&~~+{N_cg_2^2\mww\over64\pi^2}
\left[{\mtt\mbb\over\mw^4\sb^2\cb^2}\left({A_t+A_b\over\msusyy}\right)^2
 -{\mu^2\over\msusyy}\left({\mtt\over\mww\sb^2}
 +{\mbb\over\mww\cb^2}\right)^2\right]\crr
&~~-{N_cg_2^2\mtt\mbb\over192\pi^2\mww\sb^2\cb^2}
\left({A_tA_b-\mu^2\over\msusy^2}\right)^2\,,}\eqn\dmach
$$
where $(\Delta\mhpm^2)_{\rm 1LL}$ is the value of $\Delta\mhpm^2$
obtained from eq.~\llform.
This result is consistent with the one-loop calculations of
refs.~[\brignole,\marco] in the limit of large $\msusy$.
[In the same limit, the effective potential computations
of refs.~[\ellis] and [\drees] differ slightly from the above results
for the reasons  mentioned below eq.~\llform.]
Under the assumption that all the soft-supersymmetry-breaking
parameters are of the same order and
$\tanb\ll\mt/\mb$, the dominant contribution to
$\Delta\mhpm^2$ is
$$\Delta\mhpm^2 =
-{N_cg_2^2\mww\over64\pi^2}\left({\mu\over\msusy}\right)^2\left({\mt\over\sb\mw}
\right)^4
+\calo\left(g_2^2\mtt\right)\,.\eqn\dmachap$$
For sufficiently large $\mu$, this correction dominates the leading log
contributions which grow only as $\mtt$.

\chapter{Numerical Results}

In this section we evaluate the radiative corrections to the Higgs masses
and couplings. These have been computed by numerically
solving the RGEs for the Higgs self-coupling parameters to determine
the $\lambda_i(\mweak)$ as described in section 3. These results are then
inserted into eq.~\massmhh--\defghaa\
to obtain the radiatively corrected Higgs masses and couplings.
In all numerical results presented in this paper, we shall
take the squark mass parameters to be equal to a common
soft-supersymmetry breaking mass
$M_{\widetilde Q} = M_{\widetilde D} = M_{\widetilde U} = \msusy$.

\FIG\mhtbvma{%
RGE-improved Higgs mass $\mhl$ as a function
of $\tanb$ for (a) $\mt = 150$ GeV and (b) $\mt = 200$ GeV.
Various
curves correspond to $\mha = 0,~20,~50,~100$ and $300$ GeV as labeled in
the figure. All $A$-parameters and $\mu$ are set equal to zero.
The light CP-even Higgs mass varies very weakly with $\mha$
for $\mha>300$ GeV.}

In fig.~\mhtbvma (a) and (b) we
plot the light CP-even Higgs mass as a function of
$\tanb$ for $\mt = 150$ and $200$ GeV for various choices of $\mha$.
All $A$-parameters and $\mu$ are set equal to zero.
The Higgs mass saturates at a maximum value, $\mhl^{\rm max}$,
when $\tanb$ and $\mha$ become large. Furthermore, $\mhl$
converges to $\mha$ in the limit $\tanb\rarrow\infty$, as
long as $\mha\le\mhl^{\rm max}$. The reason for the $\hl$--$\ha$
mass degeneracy in this limit is easily understood. The
$\tanb\to\infty$ limit can be implemented by setting $m_{12}^2 = 0$. In
this case, the model possesses an unbroken global $\ug$ symmetry which
guarantees that $\mhl = \mha$ to all orders in perturbation theory.
That is, the radiative corrections to $\mhl$ in this
particular limit vanish exactly.\foot{If both $A\not= 0$ and
$\mu\not=0$, then the $\ug$ symmetry is not exact, and non-leading log
corrections to $\mhl$ can be generated.}
In the opposite case where $\mha\ge\mhl^{\rm max}$, $\mhh=\mha$
to all orders in perturbation theory. In contrast,
the radiative
corrections to $\mhl^2$ are substantial and grow with $\mt^4$.
Moreover, it is
easy to see that for $\mha\gg\mz$, the dominant
$\mt^4$-contribution to $\mhl^2$ is
independent of $\tanb$ [see eq.~\mhltot]. As a result,
at fixed $\mha>\mhl^{\rm max}$, the radiatively corrected $\mhl$ will
reach a maximum (minimum) at $\tanb\simeq\infty$ $(\tanb\simeq1)$,
due to the tree-level behavior of $\mhl^2$ on $\tanb$.
\FIG\regime{%
The range of allowed Higgs masses in the
large $\mha$ limit (in our plots we take $\mha=300$ GeV).
The lower limit corresponds to $\tanb=1$.
The upper limit corresponds to the limit of large
$\tanb$ (we take $\tanb=20$). In Fig.~\regime(a) and (b)
we vary $\mt$ and keep $\msusy$ fixed at $1$ and $0.5$
TeV, respectively. In Fig.~\regime(c) and (d)
we vary $\msusy$ and keep $\mt$ fixed at $150$ and $200$
GeV, respectively. The solid (dashed) curves in (c) and (d)
correspond to the computation in which the RGEs are solved numerically
(iteratively to one-loop order).}

For fixed $\tanb$, $\mhl$ reaches its minimum
value, $\mhl^{\rm min}$, when $\mha\to 0$. Note that in contrast to the
tree-level behavior (where $\mhl<\mha$), the Higgs mass does not vanish
as $\mha\to 0$. Moreover, $\mhl^{\rm min}$
increases as $\tanb$ decreases but
exhibits only a moderate dependence on $\mt$ and $\msusy$. This
behavior can be understood as follows. For $\mha\ll\mz$ and for
 values of $\mt$ and $\msusy$ sufficiently large
(say, $\mt\gsim2\mz\sb$ and $\msusy\gsim500$~GeV),
the CP-even squared mass matrix
[eq.~\massmhh] is dominated by the matrix
element $\calm_{22}^2$ due to the $\mt^4$ dependence of $\lambda_2$. This
yields
$$
(\mhl)_{\rm min}^2\simeq
\calm_{11}^2-{(\calm_{12}^2)^2\over\calm_{22}^2}\approx\mzz\cb^2
\,,\eqn\mhlll
$$
which is in good agreement with the results of
fig.~\mhtbvma. One interesting phenomenological consequence
is that $\mhl$ can be larger than $2\mha$.
This permits a new decay-mode
$\hl\rarrow\ha\ha$ which is kinematically
forbidden at tree-level. For a detailed analysis of
the $\hl\rarrow\ha\ha$ decay-mode and its implications see
refs.~[\ellis] and [\hhn].
\FIG\hlhh{%
RGE-improved
Higgs masses $\mhl$ and $\mhh$ as a
function of $\mha$, for $\msusy = 1$
TeV, $\mt = 150$ GeV and for various choices of $\tanb$.}

In the limit $\mha\to\infty$ the couplings of $\hl$ to gauge
bosons and matter fields are identical to the Higgs couplings of
the SM so that the Higgs sector of the two models cannot be
phenomenologically distinguished.
However, SUSY does impose constraints on the quartic Higgs
self-coupling at the scale $\msusy$, and this does influence
the possible values of $\mhl$. To illustrate this point, we
plot in fig.~\regime\
the range of allowed $\mhl$ in the case of large $\mha$ (in our plots we
take $\mha=300$ GeV).
As we have shown above, the lower limit for $\mhl$ is attained
if $\tanb\simeq1$ and the upper limit is attained in the limit of large
$\tanb$ (in our graphs we take $\tanb=20$).\foot{A second maximum for
$\mhl$ would arise for very
small $\tanb$; however, this regime is ruled out
because the top-Yukawa coupling develops a Landau-pole at energy scales
below $\msusy$.}
Suppose the
top quark mass is known
and that $\hl$ is discovered with SM couplings. If
$\mhl$ does not lie in the allowed mass regime displayed in fig.~\regime,
we could conclude that the MSSM is ruled out. Note that one can also
derive upper and lower Higgs mass bounds in the SM (at fixed $\mt$) as a
function of $\Lambda$. Here, $\Lambda$ is some high energy scale, below
which all Yukawa and Higgs self-coupling $\lambda$ are finite (and
$\lambda>0$ for stability of the electroweak vacuum). The lower SM Higgs
mass bound is about the same as the corresponding bound exhibited in
fig.~\regime\ for $\Lambda = \msusy$, while the upper SM Higgs mass bound
can be substantially larger than the ones exhibited in fig.~\regime\
(if $\Lambda$ is significantly smaller than the Planck scale).
The reason behind this result is the fact that in the MSSM, $\lambda$
is very small (of order $g_2^2$) at $\Lambda=\msusy$.
\FIG\contmt{%
Contours corresponding to (the radiatively
corrected) Higgs mass $\mhl$ = 20, 40, 60, 80, 100, 120 and 140 GeV
as a function of $\mt$ and $\mha$. Results are presented
for $\msusy=1$ TeV and $A_t = A_b = \mu = 0$.
We exhibit the cases of (a) $\tanb = 1$
and (b) $\tanb = 5$.}

In fig.~\hlhh(a)--(d) we plot the RGE-improved
CP-even Higgs masses $\mhl$ and $\mhh$ as functions of $\mha$ for
$\mt = 150$ GeV, $\msusy = 1$ TeV and four choices of $\tanb$.
In (a) and (b)
$\mhl$ exhibits a very weak dependence on $\mha$ as long as
$\tanb\lsim1$. In (d) with large $\tanb =
20$ we see the near mass degeneracy of $\hl$ ($\hh$)
with $\ha$ for $\mha\lsim\mz$ ($\mha\gsim\mz$) while $\mhh$ ($\mhl$)
stays constant. This behavior is due to the global $\ug$ symmetry in
the limit $m_{12}^2\to0$ as pointed out earlier.
\FIG\contur{%
Contours for constant $\mhl$
the $\tanb$--$\mha$ plane in the case $A_t = A_b = \mu = 0$.
Two adjacent graphs corresponding to $\msusy = $0.5 and 1 TeV,
and three choices of $\mt$ are displayed in both cases.
The contour labels are exhibited explicitly in
the $\msusy = 1$ TeV graphs.}

In fig.~\contmt\ we
plot contours corresponding to the radiatively corrected values of
$\mhl$ from 20 to 140 GeV in the $\mt$--$\mha$ plane.
Results are presented for $\msusy=1$ TeV and $A_t = A_b = \mu = 0$ in the
case of $\tanb = 1$ and 5.
In fig.~\contur,
contours of the radiatively corrected $\mhl$ are shown
in the $\tanb$--$\mha$
plane. Results are presented for $\msusy=0.5$ and 1 TeV and $\mt=100,150$
and $200$ GeV.

The Higgs production cross-section
in a two-Higgs-doublet model via the process
$e^+e^-\rarrow Z^*\rarrow Z\hh(Z\hl)$ is suppressed by a
factor $\cos^2(\beta-\alpha)$ [$\sin^2(\beta-\alpha)$]
as compared to the corresponding cross-sections in the SM.
\FIG\cba{%
The factor $\cos^2(\beta-\alpha)$ as a function of
$\mha$ for $\tanb = 0.5, 1 ,2$ and 20 (dotted, dashed, dot-dashed and
solid curves, respectively). Results are presented for $\msusy=1$ TeV. We
consider the case of (a) $\mt=150$ GeV and (b) $\mt=200$ GeV.}
In fig.~\cba\ we plot $\cos^2(\beta-\alpha)$ as a
function of $\mha$ for $\tanb = 0.5, 1, 2$ and 20,
for $A_t = A_b =\mu^2=0$,
$\msusy=1$ TeV and two choices of $\mt$.
The behavior is similar to that of the tree-level result in that
$\cos^2(\beta-\alpha)\to 0$ as $\mha$ becomes large. This is expected
since for large $\mha$, all heavy Higgs states decouple, while the
$\hl ZZ$ coupling [which is proportional to $\sin(\beta-\alpha)$]
approaches its SM value. Nevertheless, it is interesting to note that
$\cos^2(\beta-\alpha)$ approaches 0 more slowly as $\mt$ increases
(\ie, as the radiative corrections become more significant).

\FIG\dmnone{%
The neutral Higgs mass $\mhl$
as a function of $\tanb$ for (a)
$\mha = 50$~GeV and (b) $\mha = 300$~GeV,
for $\mt = 150$~GeV and $M_{\widetilde Q} = \msusy = 1$~TeV. All
$A$-parameters are taken to be equal $(A = A_t = A_b$); the four
contours shown correspond to $\mu = A = 0$, 1, 2 and 3 TeV,
respectively.}
\FIG\dmntwo{%
As a function of $\tanb$, we plot (a) the neutral Higgs mass $\mhl$
and (b) the factor $\sin^2(\beta-\alpha)$ for $\mt = 150$~GeV and
$M_{\widetilde Q} = \msusy = 1$~TeV. All
$A$-parameters are taken to be equal $(A = A_t = A_b$); the four
curves shown correspond to $\mu = A = 0$, 1, 2 and 3 TeV, respectively.
The sum $\mha+\mhl =\mz$ is kept fixed in both plots.}
\FIG\dmnfive{%
The range of parameters in the $\tanb$--$\mt$
plane where $\mhl = 2\mha = 40$ GeV for various choices of $\mu$ and
the $A$-parameters, and $\msusy = M_{\widetilde Q} = 1$ TeV.}
\FIG\haabrfive{%
The range of parameters in the $\tanb$--$\mha$
plane where $\mhl = 2\mha$ and
$\mt = 150$ GeV for various choices of $\mu$ and
the $A$-parameters, and $\msusy = M_{\widetilde Q} = 1$ TeV.}

Up until now, we have ignored the effects of squark mixing
by setting the $A$-parameters and $\mu$ equal to zero.
We now examine the implications of squark mixing
on the results obtained up to this point.
In order to do this, we must go beyond
the leading logarithmic approximation and include the effects of non-zero
$A$ and $\mu$ as explained in section 6.
In fig.~\dmnone\ we plot the Higgs mass $\mhl$
as a function of $\tanb$ for
$\mt = 150$ GeV and for two choices of $\mha$.
All $A$-parameters are taken to be equal; the four curves
shown correspond to $\mu = A = 0, 1, 2$ and 3 TeV, respectively.
The behavior of $\mhl$ at large values of
$\tanb$ is noteworthy: for $\mha\lsim\mz$,
we see that $\mhl$ decreases monotonically with $A$. In
contrast, in the case $\mha\gsim\mz$ and large $\tanb$,
$\mhl$ initially increases, reaches
a maximum at $A\approx \sqrt{6}\msusy$,
and then falls off rapidly\refmark\yama. These
behaviors can be obtained immediately from eqs.~\dmanl\ and \dmanh.

The results just illustrated have significant
implications for Higgs phenomenology at LEP\refmark\physlet. In
fig.~\dmntwo\ we plot the  Higgs mass $\mhl$ and the
factor $\sin^2(\beta-\alpha)$ as functions of $\tanb$
for $\mt = 150$~GeV and $\msusy = 1$~TeV. We fix the sum $\mha+\mhl
=\mz$ in both plots in order to bound $\mhl$ while keeping $Z\to\ha\hl$
kinematically forbidden. Fig.~\dmntwo(a) displays contours of fixed
$\mha+\mhl=\mz$. To the left (right) of these contours, $Z\to\ha\hl$
is kinematically allowed (forbidden).
In fig.~\dmntwo(b) we see that the decay $Z\to
Z^*\hl$, with a rate proportional to $\sin^2(\beta-\alpha)$, can be
sufficiently suppressed in the large $\tanb$ regime to escape detection.
On the other hand, the rate for $Z\to\ha\hl$ is proportional to
$\cos^2(\beta-\alpha)$ which is near unity for large $\tanb$ and
$\mha\lsim\mz$. Thus, $Z\to\ha\hl$ would be observed in this regime
unless it is kinematically forbidden. In the absence of the
Higgs discovery
at LEP, we can therefore conclude that the parameter regime to the left
of the respective curves (for various choices of $\mu = A$) shown in
fig.~\dmntwo(a) are excluded. On the other hand, at large $\tanb$,
the parameter regime to the right of the respective curves cannot
be ruled out based on current LEP data.
In particular, for large $\mu = A$ (and for large $\tanb$), the
true experimental lower limit on $\mhl$ [\ie, the dotted curve of
fig.~\dmntwo(a)] can be significantly lower than the quoted
Higgs mass limits of
the LEP detector collaborations\refmark{\pdg--\alef}.

In the search for $\ha$ at LEP via $Z\to\ha\hl$, the phenomenology
depends in detail on the decay branching ratios of $\ha$ and $\hl$.
The main impact of the one-loop radiative corrections is
a shift in the Higgs masses such that the
new (and typically dominant) decay mode
$\hl\to\ha\ha$ is now permitted.
We have already noted that the tree-level limit
$\mhl\leq\mha$ can be substantially violated when radiative corrections
are incorporated. In figs.~\dmnfive\ and \haabrfive, we depict the region
of parameter space where $\mhl\geq 2\mha$.
As before, all $A$-parameters are taken to be equal; we exhibit curves
corresponding to various choices of $A$ and $\mu$, for $M_{\widetilde
Q} = \msusy = 1$ TeV. The parameter space where $\hl\to\ha\ha$ is
kinematically allowed lies to the left of [or within] the displayed
curves.
\FIG\mucont{%
Contours of constant $\mha =0$, 40, 100 and 300 GeV in the
$\mu$--$\tanb$ plane for $\mt = 150$ GeV,
$M_{\widetilde Q} = \msusy = 1$~TeV. In (a) and (b) $\mhl = 0$
for two choices of the $A$ parameters, respectively.
In (c) and (d) $\mhl = 40$ GeV.}

In the analysis presented in this section, predictions for the
(radiatively corrected) Higgs masses are obtained as a function of
the soft SUSY breaking parameters. However, not all choices
of these parameters lead to physically (or phenomenologically)
sensible results.
For example, consider the cases of $\mt = 150$
GeV, $\msusy =1$ TeV and $\mha = 0$, 40, 100 or 300 GeV. Then the
condition $\mhl>0$ (or $\mhl> 40$ GeV if we take recent LEP limits at
face values) rules out the parameter
region above the solid, dashed or  dot-dashed
curves in fig.~\mucont.

\FIG\dpmone{%
The shift in the charged Higgs squared mass, $\Delta\mhpm^2$, normalized
to $\mww$ as a function of $\mu$ and $\tanb$ for $\mt = 2\mw$ and
$M_{\widetilde Q} = \msusy = 1$~TeV. The $A$-parameters are set to zero.
In our RGE analysis $\Delta m^2_{H^\pm}$ is independent of $\mha$.
In addition, we depict in
(a) and (b),  solid curves corresponding to $\mhl = 0$,
which depend on $\mha$ as indicated in (a).}
\FIG\dpmtwo{%
The shift in the charged Higgs squared mass, $\Delta\mhpm^2$, normalized
to $\mww$ as a function of (a) the $A$-parameter and (b) $\tanb$ for $\mt
= 2\mw$ and $M_{\widetilde Q} = \msusy = 1$~TeV. The parameter $\mu$ is
set to zero.}
\FIG\dpmthree{%
The shift in the charged Higgs squared mass, $\Delta\mhpm^2$, normalized
to $\mww$ as a function of (a) $A = A_t = A_b = \mu$
and (b) $\tanb$ for $\mt = 2\mw$ and $M_{\widetilde Q} = \msusy
= 1$~TeV.
In our RGE analysis $\Delta m^2_{H^\pm}$ is independent of $\mha$.
In addition, we depict in
(a) and (b),  solid curves corresponding to $\mhl = 0$,
which depend on $\mha$ as indicated in (a).}

Finally we consider the corrections to the charged Higgs mass sum
rule.
In figs.~\dpmone--\dpmthree\ we plot the shift in the charged Higgs
squared mass due to radiative corrections, $\Delta\mhpm^2$ [see
eq.~\defdpm],  as a function
of $\tanb$ for various choices of $A$ and $\mu$. We choose
$\mt = 2\mw$ and $M_{\widetilde Q} = \msusy =
1$~TeV.
Note that in our RGE analysis $\Delta m^2_{H^\pm}$
is independent of $m_{A^0}$.
For large values of $\tanb$ (say,
$\mb\tanb\gsim\mw$) the leading log term proportional to
$\mtt\mbb/(\mww\sb^2\cb^2)$ [see eq.~\llform] dominates and causes
$\mhpm$ to increase above its tree-level value (except in certain cases
where the effects of squark mixing become very significant).
For more moderate values of $\tanb$, a different term in  the
leading log radiative corrections, proportional to $\mt^2/\sb^2$ lowers
$\mhpm$ below its tree-level value. In addition, there is a negative
contribution to $\Delta\mhpm^2$
proportional to $\mt^4$ due to squark-mixing [see
eq.~\dmachap].  As a
result, there exists a regime in the SUSY parameter space where
$\Delta\mhpm^2<-\mww$ [see figs.~\dpmone\ and \dpmthree] which
would imply that
$\mhpm<\mha$. For example, for $\mt=2\mw$, $\tanb=1$ and $\mu =
2.5\msusy$, we find $\mhpm\approx\mha$.
Note that the radiative
corrections in the neutral Higgs sector can impose constraints
on the parameter space that depend on $\mha$ (see fig.~\mucont).
These constraints
rule out the area below and to the right of the
solid curves in figs.~\dpmone\ and
\dpmthree.

\chapter{One-Loop Leading Log vs. RGE-Improved Results}

In this section we compare the RGE-improved results (equivalent to
summing leading logs to all orders in perturbation theory) to the
one-loop leading log result
The one-loop leading log result is
obtained as described in section 3 by solving the RGEs iteratively to
first order [see eq.~\solvmzmt].
Such terms correspond precisely to a complete one-loop
perturbative calculation (with no RGE-improvement) where only the
leading log terms are retained. However, there is an ambiguity to this
procedure when it comes to diagonalizing the CP-even mass matrix
[eq.~\mthree]. If one is performing a full one-loop computation with no
RGE-improvement, then one should diagonalize this matrix perturbatively
and obtain expressions for the masses that only involve terms to first
order in the leading logs. On the other hand, a better approximation
is to diagonalize the one-loop leading log CP-even mass matrix exactly.
The resulting mass and mixing angle will differ from those
obtained by perturbative diagonalization of the mass matrix by
terms that are formally
of higher order in perturbation theory. However, this
result will be a better approximation to the full RGE-improved result,
as we shall demonstrate below.  Similar considerations have been
touched upon briefly in ref.~[\brig].

We introduce the following notation. The one-loop leading log CP-even
mass matrix [eq.~\mthree] can be diagonalized exactly to obtain the
neutral Higgs masses and mixing angle. These results will be denoted
by the abbreviation 1LL. On the other  hand, if we diagonalize
eq.~\mthree\ perturbatively and only retain the leading logs to
first order, we will use the abbreviation 1LLP. For example, from
eq.~\mthree, we find
$$
\eqalign{(\mhl^2)_{\rm 1LLP}&=
\left(\mhl^2\right)_{\rm tree} +{g_2^2m_Z^2\over96\pi^2\cw^2}
\Bigg\{\bigg[12N_c{m_b^4\sa^2\over\mz^4\cb^2}-6N_cs_{\alpha+\beta}
{m_b^2\sa\over\mzz\cb}\cr +(P_f&+P_{g}+P_{2H})(\sb^2\ca^2+\cb^2\sa^2)
-2\sb\cb\sa\ca(P_f+P_{g}'+P_{2H}') \bigg]\ln\left({\msusyy\over\mzz}
\right)\cr +&\bigg[12N_c{m_t^4\ca^2\over\mz^4\sb^2}
-6N_cs_{\alpha+\beta}{m_t^2\ca\over\mzz\sb}+s_{\alpha+\beta}^2P_t\bigg]
\ln\left({\msusyy\over m_t ^2}\right)\Bigg\}\,,}\eqn\maprox$$
where $\left(\mhl^2\right)_{\rm tree}$ is the tree-level
Higgs mass with
$\tanb$ and $\mha$ as the physical input parameters.\foot{Of course,
one has to use a definition of $\tanb$ that is consistent with $\tanb =
v_2/v_1$ at the leading log level. Eq.~\defgabb\ is an example of such
a definition.}
The CP-even mixing angle $\alpha$
and $(\mhl)_{\rm tree}$ are the tree-level values defined as:
$$
\eqalign{&\left(\tan
2\alpha\right)_{\rm tree}=\tan
2\beta\left({\mha^2+\mzz\over\mha^2-\mzz}\right)\,,\cr
&(\mhl^2)_{\rm tree}=\half(\mzz+\mha^2)\left[1+{\sin
2\beta\over(\sin2\alpha)_{\rm tree}}\right]\,.}\eqn\defta$$

\noindent
Eq.~\maprox\ plus non-leading log terms is the result one would obtain by
performing
a full one-loop calculation.

In
fig.~\regime\ we see that the 1LL results are somewhat larger than the
fully integrated RGE-improved
results. This can be understood by examining the
dominant contributions to $\beta_{\lambda_2}$. Here it is important to
note that the self-coupling constant $\lambda_2$ becomes large at low
energies and cannot be neglected. From eq.~(A.2) we
approximately have
$$
16\pi^2 (\d\lambda_2)/\d t
\simeq 6\lambda_2^2-6h_t^4+6h_t^2\lambda_2\,.\eqn\rgeapprx $$
\REF\mpi{R. Hempfling, in {\it Phenomenological
Aspects of Supersymmetry}, Lecture Notes on Physics, edited by
W. Hollik, R. R\"uckl and J. Wess (Springer-Verlag, 1992), p.~260.}
This equation has an infrared fixed point when
the right-hand side vanishes [\ie,
$h_t^2\approx\half(1+\sqrt5)\lambda_2$ or $\mhl\approx1.11\mt \sb$].
Thus the logarithmic growth of the RGE result for $\mhl$ flattens out
for large $\msusy$ as $\mhl$ approaches its fixed point value [see
fig.~\regime (c),(d)]\refmark\mpi.

\FIG\compare{%
One-loop corrected Higgs mass $\mhl$ as a function
of $\tanb$ for $\mt = 200$ GeV and 150 GeV
[(c),(d)] and $\msusy = 1$~TeV and 10~TeV. The solid
curves correspond to the fully integrated RGE-improved
Higgs mass.
The dashed curves correspond to the Higgs mass $(\mhl)_{\rm 1LL}$
derived by exactly
diagonalizing the one-loop leading log mass matrix
[eq.~\mthree]. The dotted curves correspond to $(\mhl)_{\rm 1LLP}$ given
in eq.~\maprox. In each case, the lower (upper)
curves correspond to $\mha=20$ (300)~GeV.}

In fig.~\compare\ we
plot $\mhl$ as a function of $\tanb$ for various values of $\mha$,
$\mt$ and $\msusy$.  We contrast the various methods for obtaining
the radiatively corrected value for $\mhl$. The solid curve
corresponds to the RGE-improved result obtained by numerically
solving the renormalization group equations for the $\lambda_i$
as described in section 3. The dashed curves (1LL) correspond to the
radiatively corrected $\mhl$ obtained from exactly diagonalizing the
one-loop leading log mass matrix [eq.~\mthree] and the dotted curve
(1LLP) corresponds to the one-loop perturbative result
given in eq.~\maprox. These results provide
one of the main motivations for this paper.
The comparison of the
RGE and 1LL masses shows only a small difference in the
predicted mass
in the case of $\msusy = 1$ TeV. Note that the one-loop
perturbative formula for $\mhl$ [eq.~\maprox] begins to differ
substantially from the RGE and 1LL results for low and moderate
values of $\tanb$. The comparison of 1LL and 1LLP shows agreement in
the large $\tanb$ limit as expected. Furthermore, in the large $\mha$
limit (in fig.~\compare\ we take $\mha$ = 300 GeV)
and arbitrary $\tanb$, the difference
between the 1LL and 1LLP masses is suppressed by a factor of
$\calo(\mhl^2/\mha^2)$ and is only significant for large $\mt$ and
$\msusy$. However, notice the extremely large difference between
$(\mhl)_{\rm
1LL}$ and $(\mhl)_{\rm 1LLP}$ in the case of small $\mha$ and small
$\tanb$. In particular, in contrast to the RGE and 1LL results,
$(\mhl)_{\rm 1LLP}$ is independent of $\mha$
when $\tanb = 1$.

For values of $\msusy$ much above 1 TeV [\eg, $\msusy = $10 TeV in
fig.~\compare(c) and (d)], we begin to see an appreciable
deviation between the RGE and the 1LL results when $\mha>\mz$.
In contrast, for $\mha\ll\mzz$,
the RGE and 1LL results are roughly the same. This is easily
understood---in this case,
the lightest Higgs boson is
dominantly $H_1$ and its squared mass is $\mhl^2\simeq\calm_{11}^2$.
Since
there are no large $\mt$ corrections to $\calm_{11}^2$, the
radiative corrections  to $\calm_{11}^2$ are modest and thus
exhibit very weak dependence on $\mt$ and $\msusy$. As a
result, the RGE and 1LL results are roughly the same.

To summarize, we have compared three methods for obtaining the
radiatively
corrected $\mhl$. Clearly, the RGE-improved result should be the
most complete of the three methods examined. The 1LL results
 obtained by  exactly diagonalizing eq.~\mthree\
yields results rather close to the RGE-improved
results unless $\msusy\gg 1$ TeV.\foot{For $\msusy = 1$ TeV the
largest discrepancy between these two methods occurs for large
$\mt$ and $\mha\gg\mz$. For example, for $\tanb = 1$, $\mha = 300$
GeV and $\mt = 200$ GeV, we find $(\mhl)_{\rm RGE} = 96.8$ GeV
while $(\mhl)_{\rm 1LL} = 104.4$ GeV. In contrast, for the same set of
parameters we find $(\mhl)_{\rm 1LLP} = 111.5$
GeV.}

Of the three methods for computing the Higgs mass discussed
in this section, the fully integrated RGE analysis provides the
best approximation.  How accurate will such an approximation be?
To fully answer this question requires a detailed
investigation of the non-leading logarithmic terms that have been
neglected in our analysis.
In this paper, we have identified the most important
terms of this type only in the limit of large squark mixing. To
make further progress requires a complete one-loop
computation\refmark{\leter,\pokor,\diaz}.
The most
accurate values for the radiatively corrected Higgs mass would
then be obtained by identifying the complete set of non-leading
logarithmic terms (say by subtracting the results of
eqs.~\mthree--\llform\ from the corresponding exact one-loop
calculation) and adding these to our fully-integrated
RGE results. By comparing our results to those of a full
one-loop calculation, we conclude that the effect of the
non-leading corrections on the neutral CP-even Higgs masses is
typically
no more than a few GeV.
Clearly, the relative importance of such terms becomes less significant
as $\msusy$ increases. We can improve our results slightly by identifying
the largest of the non-leading logarithmic corrections not yet included.
In the case of the CP-even Higgs mass-squared matrix, when $\mt\gg\mw$,
the largest of such terms are ones
of $\calo(g_2^2\mtt)$.
{}From a full one-loop perturbative computation,
$$
\calm^2 = \calm^2_{\rm RGE} +
{N_cg_2^2\mtt\over48\pi^2\sb^2c_W^2}\left(\matrix{0&0\cr0&1}\right)
\eqn\nllogtrm
$$
where $\calm^2_{\rm RGE}$ is the RGE-improved result [\ie,
eq.~\massmhh\ with the $\lambda_i$ obtained by numerically solving the
RGEs). The shift of  the
light Higgs mass due to these non-leading log corrections is of the
order of 1 GeV.
The largest non-leading log corrections to $\mhpm^2$ are given in
ref.~[\marco].
With this final improvement, our formulae for radiatively
corrected Higgs masses are quite accurate in nearly all regions
of SUSY parameter space for $\msusy\gsim 400$ GeV.

\chapter{Conclusions}

We have calculated the dominant radiative corrections to the Higgs
sector parameters in the MSSM. We have obtained analytic formulae for
the one-loop leading logarithmic corrections as a function of the
various supersymmetric parameters. Our analysis also includes the
most important non-logarithmic effects due to squark mixing.
In our numerical analysis, we have focused on the case where
all the supersymmetric partners are approximately mass degenerate.
Summation
of the leading logarithms
to all orders in perturbation theory
is achieved
by solving the RGEs numerically. Non-leading logarithmic corrections
due to squark mixing effects are included by computing the Higgs
4-point functions at the scale $\msusy$ and modifying the
supersymmetric boundary condition.

\REF\radpheno{R. Barbieri and M. Frigeni, {\sl Phys. Lett.} {\bf B258},
395 (1991); J.L. Lopez
and D.V. Nanopoulos, {\sl Phys. Lett.} {\bf B266}, 397 (1991).}
The most significant contribution to
the radiative corrections to $\mhl$ grows with
$\mt^4$ and logarithmically with $\msusy$.
A number of the important phenomenological implications of the Higgs
radiative corrections
have already been obtained elsewhere (see \eg, refs.~%
[\erz,\ellis, and \radpheno]).
The MSSM cannot be ruled
out if LEP-200 fails to discover the Higgs boson with a mass
$\mhl\lsim\mz$. In particular, if $\mt\gsim 150$ GeV, then
$\mhl\gsim\mz$ for large $\mha$ and $\tanb$ if $\msusy$ is sufficiently
above $\mt$. For small values of $\mha$ ($\lsim 30$ GeV) and small
values of $\tanb$ ($\lsim 1$) the light Higgs mass can be
larger then $2\mha$. In this case the decay $\hl\to\ha\ha$ is
kinematically allowed and provides the dominant decay mode over a
significant region of the parameter space. The leading log corrections to
$\mhpm$ only grow with $\mt^2$. However, there are non-logarithmic
contributions due to squark mixing effects that grow as $\mt^4$. These
effects could (in extreme cases) yield
$\mhpm\lsim\mha$. Large squark mixing effects  can also lower the
experimental lower bounds on $\mhl$ in the large $\tanb$ region.

Numerical comparison of the first order leading log corrections
obtained in this paper with the full one-loop radiative corrections
in the limit of large $\tanb$\refmark{\leter,\pokor,\diaz}\
shows agreement within a few
percent or better for
$\msusy \gsim 400$ GeV. The non-leading log terms not included in our
analysis are thus of the same order as higher order leading log
terms which are summed by renormalization group
improvement and both should be included in any full one-loop
calculation. The one-loop leading log expressions also serve
as a useful check of any complete one-loop computation.
\bigskip

\centerline{{\bf Acknowledgments}}
\smallskip
We gratefully acknowledge discussions with R. Barbieri, M. D\'\i az,
J. Ellis, Y. Okada, K. Sasaki, A. Yamada, and F. Zwirner.  This work
was supported in part by the U.S. Department of Energy.

\endpage

\Appendix{A}
\medskip
\centerline{\bf Renormalization Group Equations}
\medskip

\REF\rge{K. Inoue, A. Kakuto and Y. Nakano, Prog. Theor.
Phys. {\bf 63}, 234 (1980); H. Komatsu, {\sl Prog. Theor. Phys.}
{\bf 67}, 1177 (1982); C.T. Hill, C.N. Leung and S. Rao,
{\sl Nucl. Phys.} {\bf B262}, 517 (1985).}
\REF\rgesusy{K. Inoue, A. Kakuto, H. Komatsu, and  S. Takeshita,
{\sl Prog. Theor. Phys.} {\bf 67}, 1889 (1982); {\bf 68}, 927 (1982)
[E: {\bf 70}, 330 (1983)]; {\bf 71}, 413 (1984); B. Gato, J. Le\'on,
J. P\'erez-Mercader and M. Quir\'os, {\sl Nucl. Phys.} {\bf B253}, 285
(1985); N.K. Falck, {\sl Z. Phys.} {\bf C30}, 247 (1986).}
\REF\rgedec{Supersymmetric RGEs in the limit
where squarks and gluinos are decoupled have been given previously
in P.H. Chankowski, {\sl Phys. Rev.} {\bf D41}, 2877 (1990).}
The
$\beta$-functions in the non-supersymmetric
two-Higgs-doublet model with the discrete
symmetry $\Phi_1\rarrow-\Phi_1$ and in the MSSM
are well known\refmark{\chengli,\bagger,\rge,\rgesusy}.
Here, we generalize these results
in two respects. First, we take the most general (CP-conserving)
 two-Higgs-doublet potential
[\ie, with no discrete symmetry] given in eq.~\pot.
Second, we explicitly treat supersymmetric particle
contributions to the $\beta$-functions in the step-approximation.
That is, a particular SUSY contribution is omitted for scales below the
corresponding SUSY particle mass\refmark\rgedec.  We then find
$$\eqalign{
16&\pi^2\beta_{\lambda_{1}} = \left\{6\lambda^2_{1}+\lambda_3^2
+\left(\lambda_3+\lambda_4\right)^2
+\lambda_5^2+12\lambda_{6}^2+\threeighth
[2g_2^4+(g_2^2+g_1^2)^2]\right\}\theta_Z\cr
&+\sum_i N_{c_i}\left\{-2h_{D_i}^4 \theta_Z
 +\left(h_{D_i}^2-\fourth
g_1^2Y_{D_i}\right)^2\theta_{\widetilde D_i}
 +\left(\fourth
g_1^2Y_{U_i}\right)^2\theta_{\widetilde U_i}\right.\cr
&\qquad\left.+\left[h_{D_i}^4-\half h_{D_i}^2(g_1^2Y_{Q_i}+g_2^2)
+\eighth(g_2^4+g_1^4Y_{Q_i}^2)\right]\theta_{\widetilde
Q_i}\right\}\cr
&-\fivehalf
 g_2^4\theta_{\widetilde
H}\theta_{\widetilde W}-g_1^2g_2^2\theta_{\widetilde
H}\theta_{\widetilde W}\theta_{\widetilde B}
-\half g_1^4\theta_{\widetilde
H}\theta_{\widetilde B}-32\pi^2\lambda_{1}\gamma_{1}\,,}\eqn\betalone
$$
$$\eqalign{
16&\pi^2\beta_{\lambda_{2}}=\left\{6\lambda^2_2+\lambda_3^2
+\left(\lambda_3+\lambda_4\right)^2
+\lambda_5^2+12\lambda_{6}^2+\threeighth
[2g_2^4+(g_2^2+g_1^2)^2]\right\}\theta_Z\cr
&+\sum_i
N_{c_i}\left\{-2h_{U_i}^4\theta_{U_i}\theta_Z +\left(\fourth
g_1^2Y_{D_i}\right)^2\theta_{\widetilde D_i}
 +\left(h_{U_i}^2+\fourth
g_1^2Y_{U_i}\right)^2\theta_{\widetilde U_i}\right.\cr
&\qquad\left.+\left[h_{U_i}^4+\half h_{U_i}^2(g_1^2Y_{Q_i}-g_2^2)
+\eighth(g_2^4+g_1^4Y_{Q_i}^2)\right]\theta_{\widetilde Q_i}\right\}\cr
&-\fivehalf
 g_2^4\theta_{\widetilde
H}\theta_{\widetilde W}-g_1^2g_2^2\theta_{\widetilde
H}\theta_{\widetilde W}\theta_{\widetilde B}
-\half g_1^4\theta_{\widetilde
H}\theta_{\widetilde B}-32\pi^2\lambda_2\gamma_2\,,}\eqn\betaltwo
$$
$$\eqalign{
16&\pi^2\beta_{\lambda_3}=\left\{(\lambda_1
+\lambda_2)(3\lambda_3+\lambda_4)+2\lambda_3^2
+\lambda_4^2+\lambda_5^2+2\lambda_6^2+2\lambda_7^2
+8\lambda_6\lambda_7\right.\cr
&\qquad\left.+\threeighth
\big[2g_2^4+(g_2^2-g_1^2)^2\big]\right\}\theta_Z\cr
&+\sum_i N_{c_i}\left\{-2h_{U_i}^2 h_{D_i}^2
\theta_{U_i}\theta_Z
+\fourth g_1^2Y_{D_i}\left(h_{D_i}^2-\fourth
g_1^2Y_{D_i}\right)\theta_{\widetilde D_i}\right. \cr
&\qquad-\fourth g_1^2Y_{U_i}\left(h_{U_i}^2+\fourth
g_1^2Y_{U_i}\right)\theta_{\widetilde U_i}+h_{U_i}^2h_{D_i}^2
\theta_{\widetilde U_i}\theta_{\widetilde D_i}
\cr
&\qquad\left.+\left[h_{U_i}^2h_{D_i}^2-\fourth
h_{U_i}^2(g_1^2Y_{Q_i}+g_2^2) +\fourth
h_{D_i}^2(g_1^2Y_{Q_i}-g_2^2)+\eighth(g_2^4-g_1^4Y_{Q_i}^2)\right]
\theta_{\widetilde Q_i}\right\}\cr
&-\fivehalf
g_2^4\theta_{\widetilde
W}\theta_{\widetilde H}+g_1^2g_2^2\theta_{\widetilde
W}\theta_{\widetilde B}\theta_{\widetilde H} -\half
g_1^4\theta_{\widetilde B}\theta_{\widetilde H}
-16\pi^2\lambda_3(\gamma_1+\gamma_2)\,,}\eqn\betalthree
$$
$$\eqalign{
16&\pi^2\beta_{\lambda_4}=\left[ \lambda_4 \right. (\lambda_1
+\lambda_2+4\lambda_3+2\lambda_4)+              \left.
4\lambda_5^2+5\lambda_6^2+5\lambda_7^2+2\lambda_6\lambda_7+\threehalf
g_2^2g_1^2\right]\theta_Z\cr
 &+\sum_i N_{c_i}\left\{
2h_{U_i}^2h_{D_i}^2\theta_{U_i}\theta_Z
-h_{U_i}^2h_{D_i}^2
\theta_{\widetilde U_i}\theta_{\widetilde D_i}
-(h_{U_i}^2-\half
g_2^2)(h_{D_i}^2-\half g_2^2) \theta_{\widetilde
Q_i}\right\}\cr
 &+2g_2^4\theta_{\widetilde
W}\theta_{\widetilde H}-2g_1^2g_2^2\theta_{\widetilde
W}\theta_{\widetilde B}\theta_{\widetilde H}
 -16\pi^2\lambda_4(\gamma_1+\gamma_2)\,,}\eqn\betalfour
$$
$$\eqalign{
16\pi^2\beta_{\lambda_5} =
\left[  \lambda_5 \right.&(\lambda_1+\lambda_2+4\lambda_3+6\lambda_4)
+5\big(\lambda_6^2+\lambda_7^2\big)  \left.
+2\lambda_6\lambda_7\right]\theta_Z\cr
&-16\pi^2\lambda_5(\gamma_1+\gamma_2)\,,}\eqn\betalfive
$$
$$\eqalign{
16\pi^2\beta_{\lambda_6} =
\left[ \lambda_6\right. &\big(6\lambda_1+3\lambda_3
+4\lambda_4+5\lambda_5\big) +\lambda_7 \big(\left.
3\lambda_3+2\lambda_4+\lambda_5\big)\right]\theta_Z\cr
&-8\pi^2\lambda_6(3\gamma_1+\gamma_2)\,,}\eqn\betalsix
$$
$$\eqalign{
16\pi^2\beta_{\lambda_7} =
\left[ \lambda_7 \right. &\big(6\lambda_2+3\lambda_3+4\lambda_4
+5\lambda_5\big) +\lambda_6\big(  \left.
3\lambda_3+2\lambda_4+\lambda_5\big)\right]\theta_Z\cr
&-8\pi^2\lambda_7(\gamma_1+3\gamma_2)\,;\cr
}\eqn\betalseven$$
where
$$
\theta_X\equiv\theta\left(\ln{s\over M_X^2}\right) =
\cases{1& for $s \ge M_X^2$,\cr
       0& for $s < M_X^2$,\cr}
\eqn\deftheta$$
and $M_X$ is the mass of the field $X$.
To first order in $\mweak^2/\msusy^2$, we can ignore mixing in the
supersymmetric mass matrices.
In this case, the supersymmetric particle spectrum consists
of the gluino ($\tilde g$), neutralinos ($\widetilde B,\widetilde W^3,
\widetilde H^0_1,\widetilde H^0_2$),
charginos ($\widetilde W^\pm, \widetilde H^\pm$),
L-type squarks and sleptons
($\widetilde Q^1_i,\widetilde Q^2_i$), and R-type squarks and
sleptons ($\widetilde U_i,\widetilde D_i$).  The corresponding
supersymmetric particle masses are given by:
$M_{\widetilde g}=\abs{M_3}$,
$M_{\widetilde B}=\abs{M_1}$, $M_{\widetilde W}=\abs{M_2}$,
$M_{\widetilde H}=\abs{\mu}$, the L-type squarks [sleptons] are
degenerate with mass $M_{\widetilde Q_i}$ and the R-type squark
[slepton] masses are $M_{\widetilde U_i}$ and $M_{\widetilde D_i}$.
The label $i$ runs over three generations and counts both squarks
and sleptons; the corresponding fermion partners (\ie,
quarks and leptons) are denoted by $U_i$ and $D_i$.
The up- and down-type Yukawa couplings are
proportional to the corresponding quark [lepton] masses
$$
h_{U_i}={gm_{U_i}\over\sqrt2 m_W\sinb}\,,\qquad
h_{D_i}={gm_{D_i}\over\sqrt2 m_W\cosb}\,.\eqn\defyukawa
$$
The color factor $N_{c_i} = 3$ [1] when the label $i$ refers
to (s)quarks [(s)leptons],
and the hypercharges
of the squarks [sleptons] are $Y_{Q_i}=\third$ [$-1$],
$Y_{D_i}=\twothirds$ [2] and $Y_{U_i}=-\fourthirds$
[there is no $\widetilde\nu_R$].
Finally, the anomalous
dimensions $\gamma_j$ ($j = 1, 2$) in the Landau gauge
are given by
$$\eqalign{
\gamma_1={\d\over \d t} \ln{\Phi_{1}^2}={1\over 64\pi^2}
&\left[\left(9g_2^2+3g_1^2-4\sum_i
N_{c_i}h_{D_i}^2\right)\theta_Z\right.\crr
&~\left.-6g_2^2\theta_{\widetilde W}\theta_{\widetilde H}
-2g_1^2\theta_{\widetilde B}\theta_{\widetilde H}\right]\,,\crr
\gamma_2={\d\over \d t}\ln{\Phi_{2}^2}={1\over
64\pi^2}&\left[\left(9g_2^2+3g_1^2-4\sum_i
N_{c_i}h_{U_i}^2\theta_{U_i}\right)\theta_Z\right.\cr
&~\left.-6g_2^2\theta_{\widetilde W}\theta_{\widetilde
H}-2g_1^2\theta_{\widetilde B}\theta_{\widetilde H}\right]\,.
}\eqn\wavez$$

The $\beta$-functions for the $\uy$, $\sutwol$ and $\suc$ gauge
couplings $g_1\equiv g'$ $g_2\equiv g$ and $g_3\equiv g_s$ are
$$\eqalign{
48\pi^2\beta_{g_1^2}&=\biggl(\fourth\sum_i
N_{c_i}\left[
2Y_{Q_i}^2\left(2\theta_t+\theta_{\widetilde Q_i}\right)
+Y_{U_i}^2\left(2\theta_t+\theta_{\widetilde U_i}\right)
+Y_{D_i}^2\left(2\theta_t+\theta_{\widetilde D_i}\right)\right]\cr
&\qquad+\half N_H\theta_t+N_{\tilde H}\theta_{\widetilde H}\biggr)
g_1^4\,,\cr
48\pi^2\beta_{g_2^2}&=\left(\half\sum_i N_{c_i}
\left(2\theta_t+\theta_{\widetilde Q_i}\right)
+\half N_H\theta_t+N_{\tilde H}\theta_{\widetilde H}
+4N_{\tilde w}\theta_{\widetilde W}-22\theta_t\right)g_2^4\,,\cr
48\pi^2\beta_{g_3^2}&=\left(\sum_i c_i
\left( 2\theta_{U_i}+2\theta_{D_i}
+\theta_{\widetilde Q_i}
+\half\theta_{\widetilde U_i}
+\half\theta_{\widetilde D_i}\right)
+6N_{\tilde o}\theta_{\tilde g}-33\right)g_3^4\,,\cr
}\eqn\betagg$$
where $c_i = 1$ [0] for color triplet [singlet] fermions and
their supersymmetric partners.
The number of Higgs doublets is $N_H = 2$,
the number of higgsino doublets is $N_{\tilde H}=2$, the
number of fermionic left-handed triplets is
$N_{\tilde w}=1$ (the $\widetilde W$) and the
number of fermionic color octets is $N_{\tilde o}=1$
(the gluino $\tilde g$).
The $\beta$-functions for $g_1$ and $g_2$
are valid for scales $\sqrt{s}>\mt$ where
the electroweak gauge theory is unbroken;
for the $\beta$-functions below
top quark-threshold see Appendix B.
The boundary
conditions are $g_k^2(\mzz)= 4\pi/(128 c_W^2),\ 4\pi/(128s_W^2)$ and
1.38 for $k = 1, 2, 3$, respectively. We take $s_W^2 = 0.23$ in our
numerical work.
One can explicitly check that
eq.~\bndbeta\ is satisfied for scales larger than the largest SUSY
mass parameter.

It is important to clarify the validity of the above formulae
with explicit $\theta$-function treatment of thresholds.
If all supersymmetric particle masses are roughly degenerate
and $\mt\simeq\mz$, then integration of the above formulae
will correctly sum all leading logarithmic contributions
to all orders in perturbation theory. However, if there are
non-degenerate thresholds this is no longer the case.
For example, the treatment of the top quark threshold is discussed in
Appendix B. Below this threshold
the electroweak gauge group is broken and thus the effective
low-energy theory now has many new terms which require the
introduction of new couplings. One could work out the complete set of
RGEs, including those for all couplings (with appropriate
boundary conditions at $\sqrt{s}=\mt$ corresponding to
the requirement of electroweak gauge symmetry). In practice, such a
procedure would be overkill; after all, $\mt$ is not much larger
than $\mz$. In Appendix B we show how to correctly obtain the
one-loop leading logarithmic terms proportional to
$g_2^2\ln(\mt^2/\mzz)$. These are certainly sufficient for our
purposes.

A similar set of remarks hold for the supersymmetric
particle thresholds.\foot{In the numerical work presented in this
paper, all supersymmetric masses are taken to be degenerate.
Thus the considerations below do not affect these results.}
Let $\msusy^{\rm max}$ be the largest supersymmetry-breaking mass
parameter in the MSSM. Then for $\sqrt{s} < \msusy^{\rm max}$,
we should in principle consider a new low-energy effective theory
with vertices that are no longer constrained by SUSY. This
would introduce many new couplings, and the RGEs for these couplings
would be required in order to sum logarithmic terms such as
$\ln(M_i^2/M_j^2)$ to all orders. ($M_i$ and $M_j$ are masses of
different supersymmetric particles.) Again, this is much
more than we need. In deriving the RGEs above we have implicitly assumed
supersymmetric relations among the various couplings of the theory
for all $\msusy^{\rm min}<\sqrt{s}<\msusy^{\rm max}$
(where we have assumed that the mass of the lightest
supersymmetric particle, $\msusy^{\rm min}$ lies above $\mz$).
Integrating the RGEs listed above will then yield terms of
order $g_2^2\ln(M_i^2/M_j^2)$, although these terms
will not be correctly summed to all orders.\foot{%
Such terms are typically not as significant as the terms proportional to
$g^2_2 \ln(m^2_t/m^2_Z$), which appear multiplied by powers
of $m_t/m_Z$ in the formulae for Higgs masses and couplings.}
On the other hand, the set of RGEs corresponding
to the (non-supersymmetric) two-Higgs-doublet model
({\it i.e.}, for $\sqrt{s}\leq\msusy^{\rm min}$) is sufficient
to sum to all orders logarithmic terms proportional to
$\ln(\msusy^2/\mweak^2$), where $\msusy$ is some average
supersymmetry breaking mass.
Results obtained in this manner are certainly accurate
enough for our purposes.

Thus, to complete the set of RGEs required for the computation of the
$\lambda_i(\mz)$, we need RGEs for the Yukawa couplings corresponding
to the (non-supersymmetric) two-Higgs-doublet
model\refmark{\chengli,\bagger,\rge}.
These are given by
$$
\eqalign{
16\pi^2\beta_{h_t^2}&=\left(\ninehalf
h_t^2+\half h_b^2-8g_3^2-\ninefourth g_2^2-\seventeentwelfth
g_1^2\right)h_t^2\,,\cr
16\pi^2\beta_{h_b^2}&=\left(\ninehalf
h_b^2+\half h_t^2+h_{\tau}^2-8g_3^2-\ninefourth
g_2^2-\fivetwelfth g_1^2\right)h_b^2\,,\cr
16\pi^2\beta_{h_{\tau}^2}&=\left(\fivehalf
h_{\tau}^2+3h_b^2-\ninefourth g_2^2-\fifteenfourth
g_1^2\right)h_{\tau}^2\,.\cr
}\eqn\rgeh$$
All other Yukawa couplings can be neglected.
Note that eq.~\rgeh\ assumes that the Higgs--fermion interactions have
the same structure as the ones in the MSSM.
For completeness we present the
$\beta$-functions for the mass parameters (although these are not
required for any of the applications presented in this paper)
$$\eqalign{&16\pi^2\beta_{m_{11}^2}=
\big[3m_{11}^2\lambda_1+m^2_{22}(2\lambda_3+\lambda_4)
-6m_{12}^2\lambda_6\big] -\gamma_{1}m_{11}^2\cr
&16\pi^2\beta_{m_{22}^2}=
\big[3m_{22}^2\lambda_2+m^2_{11}(2\lambda_3+\lambda_4)
-6m_{12}^2\lambda_7\big] -\gamma_{2}m_{22}^2\cr
&16\pi^2\beta_{m_{12}^2} =
\Big[(\lambda_3+2\lambda_4+3\lambda_5)m^2_{12}
-3\lambda_6m_{11}^2-3\lambda_7m_{22}^2\Big]-\half(\gamma_1+\gamma_2)m^2_{12}\,.\
cr}
\eqn\betam$$

\Appendix{B}
\medskip
\centerline{\bf Coupling Constant Evolution Below the Top Quark
Threshold}
\medskip

In this appendix we present the $\beta$-functions for the relevant
gauge and Higgs self-interaction coupling constants below
top quark-threshold. Note that by formally integrating out the
top-quark, we include logarithmic contributions to the
Higgs mass matrix of $\calo[g_2^2\mzz\ln(\mtt/\mzz)]$. However,
it is important to emphasize that such terms are formally
smaller or of similar size to
some non-leading log terms, \eg, terms of
$\calo(g_2^2\mtt$). [The latter can be included easily according to
eq.~\nllogtrm.]
Nevertheless, isolating
such logarithmic terms serves as a useful check of more complete
one-loop computations.

First consider the $\beta$-functions for the Higgs self-couplings.
These determine the running of the $\lambda_i$; the values of
$\lambda_i(\mweak)$ enter the calculations of the neutral and
charged Higgs parameters. In calculations involving the neutral Higgs
sector the contributions from the top quark can be removed by setting
$h_t=0$. In the charged sector we note that some (but not all)
of the pieces
proportional to $h_b^2$ arise from $t$--$b$ loops and should also
be removed.
For the purpose of the
calculations presented here, the simpler procedure of
setting $h_t=0$ for the neutral Higgs processes and $h_t=h_b=0$ for
charged Higgs processes suffices.\foot{In a more precise procedure
one would have to write down the most general potential for
four neutral fields and four charged fields invariant under $\uem$
and determine the bottom quark contributions to the $\beta$-function
for each coupling constant separately. The boundary conditions for
these coupling constants are obtained by requiring that the potential
reduces to eq.~\pot\ at a scale $\sqrt{s}=\mt$.}

Next we look at the $\beta$-functions of the gauge coupling constants.
First we note that below the top quark-threshold
the $\sutwouone$ gauge symmetry is broken down to $\uem$. The
coupling constants of the photon to charged particles are
constrained by $\uem$ gauge invariance. In contrast, the
couplings of the $W$ and $Z$ are now independent parameters
which thus can evolve differently from $\mt$ to $\mz$. To
determine the evolution of the gauge couplings, we
fix as experimental inputs the masses of the $W$ and $Z$
bosons $\mz$ and $\mw$ which arise through the
interaction
$$
\call = \fourth\left(\half
G^{ij}_{kl}V_{\mu i}V^\mu_j+ G_{kl}^\pm W_\mu^+W^{\mu
-}\right)H^0_kH^0_l\,,\eqn\ggpot
$$
where $V_i = (W^3, B)$ are the neutral $\sutwol$ and $\uy$
gauge fields.  At the scale $\mt$ where the $\sutwouone$ gauge
symmetry is restored we impose the boundary conditions
$$
\eqalign{&G_{kl}^\pm=\delta_{kl}g_2^2\,,\cr
&G^{ij}_{kl}=\delta_{kl}\bar g_i\bar g_j\qquad\hbox{with}\qquad
\bar g_i=(g_1,-g_2)\,.}\eqn\bndgg
$$
In general the
$\beta$-functions corresponding to the four-point interactions,
eq.~\ggpot, have two types of contributions: the box diagrams
and
the wave function renormalizations. In the following we consider
only diagrams involving the top-quark. It is clear that the
quark loop contributions to the wave function renormalization
of the Higgs boson and the box diagrams to $G_{kl}^{ij}$
cancel since the $\beta_{g_i^2}$ in eq.~\betagg\ do not contain
pieces proportional to $h_t^2$ or $h_b^2$. Below $\mt$ we
remove the $H^-tb$ interaction from the theory so that there are no
vertex corrections to $G_{kl}^\pm$ from the bottom quark.
However, there are self energy diagrams to both the gauge bosons
and the Higgs bosons due to the bottom quark. In the regime where
$\tanb\ll\mw/\mb$ (\ie, $h_b\ll g_2$) we can neglect the Higgs boson
self energy diagram. Thus the only other diagrams one has to
consider are the self energy diagrams for the gauge bosons. We
assert that
$$\eqalign{
&G_{kl}^\pm=G^\pm\delta_{kl}\,,\cr
&G_{kl}^{ij}=G^{ij}\delta_{kl}\,.
}\eqn\ansz$$
Then the RGEs of
$G^{ij}$ and $G^\pm$ differ from the RGEs of the gauge coupling
only by the top-quark contributions
$$\eqalign{
&\beta_{G^\pm}=\beta_{g_{2}^2} +G^\pm\gamma_\pm^t\,,\cr
&\beta_{G^{ij}}=\beta_{g_{i}^2}
\delta_{ij}+\half\big(G^{ik}\gamma_{kj}^t
+G^{jk}\gamma_{ki}^t\big)\,,
}\eqn\rgegg$$
(no summation over $i$). The anomalous dimensions $\gamma_{ij}$
and $\gamma_\pm$ are defined as
$$\eqalign{ \gamma_\pm
\equiv&{\d\,\ln (W_{\mu}^+W^{-\mu})\over\d t} ={\d{\rm
A}_{WW}'(p^2)\over\d t} \,,\cr
\gamma_{ij}\equiv&{\d\,\ln
(V\ls{\mu i}V^{\mu}_j)\over\d t} ={\d{\rm
A}_{V_iV_j}'(p^2)\over\d t} \,,
}\eqn\defwave$$
where ${\rm
A}_{VV}$ $(V=V_i,W)$ are proportional to the
$g_{\mu\nu}$ term of the self-energies of the gauge bosons and
${\rm A}_{VV}'(p^2)\equiv\d {\rm A}_{VV}/\d p^2$. The
contributions involving the top-quark are
$$\eqalign{&\gamma_{\pm}^t=-{g_2^2\over16\pi^2}\,,\cr
&\gamma_{ij}^t=-{1\over32\pi^2}\big(a_i^La_j^L+a_i^Ra_j^R\big)
\,,}\eqn\waverge$$
where $a_i^L=(g_2,\third g_1)$ and
$a_i^R=(0,\fourthirds g_1)$.

\vskip1cm
\Appendix{C}
\medskip
\centerline{\bf One-Loop Leading Logarithmic Higgs Self-Couplings}
\medskip
In the analysis presented in this paper, the RGEs of Appendix A are
solved numerically for $\mt<\sqrt{s}<\msusy$ and iteratively to
one-loop order for $\mweak<\sqrt{s}<\mt$ using the RGEs described
in Appendix B.
However, it is instructive to solve the RGEs completely
iteratively to one-loop order [see eq.~\solvmzmt].
In obtaining the results below, we have carefully removed the
top-quark from the low-energy effective theory at energy scales
below $m_t$ as explained in section 3.  We find
$$\eqalign{
\lambda_1&(\mz) = \fourth[g_1^2+g_2^2](\mz)
   +{g_2^4\over384\pi^2\cw^4}\left\{\vbox to 21pt{}
   6\sww\left(1-2\sww\right)t_Z^{\widetilde\chi_1}
   -24\sww\left(1-\sww\right)t_Z^{\widetilde\chi_{12}}\right.\crr
  &-\left(42-102\sww+60\sw^4\right)t_Z^{\widetilde\chi_2}
   -4\left(1-2\sww+2\sw^4\right)t_Z^{\widetilde H}
   -8\left(1-2\sww+\sw^4\right)t_Z^{\widetilde W}\crr
   &+\sum_i N_{c_i}\left[\left(6
   {m_{D_i}^4\over\mz^4\cb^4}+12{m_{D_i}^2\over\mz^2\cb^2}e_{D_i}\sww
    +4e_{D_i}^2\sw^4\right)t_Z^{\widetilde D_i}\right.\crr
  &+\left(6{m_{D_i}^4\over\mz^4\cb^4}
    -6{m_{D_i}^2\over\mz^2\cb^2}
    \left[ \cww+\sww(e_{U_i}+e_{D_i})\right]
    +1+4e_{D_i}\sww+4e_{D_i}^2\sw^4\right)t_Z^{\widetilde Q_i}\crr
  &\left.\left.+4e_{U_i}^2\sw^4
t_{U_i}^{\widetilde U_i}+\left(1-4e_{U_i}\sww+4e_{U_i}^2\sw^4
     \right)t_{U_i}^{\widetilde Q_i}\vbox to 21pt{}\right]
     \right\}\,,
}\eqn\blabla$$
$$\eqalign{
\lambda_2&(\mz) = \fourth
   [g_1^2+g_2^2](\mz)
   +{g_2^4\over384\pi^2\cw^4}\left\{ \vbox to 21pt{}
  6\sww\left(1-2\sww\right)t_Z^{\widetilde\chi_1}
-24\sww\left(1-\sww\right)t_Z^{\widetilde\chi_{12}}\right.\crr
 &-\left(42-102\sww+60\sw^4\right)t_Z^{\widetilde\chi_2}
   -4\left(1-2\sww+2\sw^4\right)t_Z^{\widetilde H}
   -8\left(1-2\sww+\sw^4\right)t_Z^{\widetilde W}\crr
   &+\sum_i N_{c_i}\left[\left(6
   {m_{U_i}^4\over\mz^4\sb^4}-12{m_{U_i}^2\over\mz^2\sb^2}e_{U_i}\sww
   +4e_{U_i}^2\sw^4\right)t_{U_i}^{\widetilde U_i}\right.\crr
 &+\left(6{m_{U_i}^4\over\mz^4\sb^4}
   -6{m_{U_i}^2\over\mz^2\sb^2}
   \left[\cww-\sww(e_{U_i}+e_{D_i})\right]
   +1-4e_{U_i}\sww+4e_{U_i}^2\sw^4\right)t_{U_i}^{\widetilde Q_i}\crr
 &\left.\left.+4e_{D_i}^2\sw^4
   t_Z^{\widetilde D_i}+\left(1+4e_{D_i}\sww
  +4e_{D_i}^2\sw^4\right)t_Z^{\widetilde Q_i}
    \vbox to 21pt{}\right] \right\}\,,
}\eqn\blabla$$
$$\eqalign{
\widetilde\lambda_3&(\mz)=-\fourth[g_1^2+g_2^2](\mz)
    -{g_2^4\over384\pi^2\cw^4}\left\{ \vbox to 21pt{}
    6\sww\left(1+2\sww\right)t_Z^{\widetilde\chi_1}
   +24\sww\left(1-\sww\right)t_Z^{\widetilde\chi_{12}}\right.\cr
   &+\left(30-42\sww+12\sw^4\right)t_Z^{\widetilde\chi_2}
   -4\left(1-2\sww+2\sw^4\right)t_Z^{\widetilde H}
   -8\left(1-2\sww+\sw^4\right)t_Z^{\widetilde W} \cr
   &+\sum_i N_{c_i}\left[\left(-6{m_{U_i}^2\over\mzz\sb^2}+4\sww
    e_{U_i}\right)\sww e_{U_i}
    t_{U_i}^{\widetilde
U_i}+\left(6{m_{D_i}^2\over\mzz\cb^2}+4\sw^2
    e_{D_i}\right)\sww e_{D_i} t_Z^{\widetilde D_i}\right.\cr
 &+\left(-3{m_{U_i}^2\over\mz^2\sb^2}
   \left[\cww-\sww(e_{U_i}+e_{D_i})\right]
   +1-4e_{U_i}\sww+4e_{U_i}^2\sw^4\right)t_{U_i}^{\widetilde Q_i}\cr
 &+\left.\left.\left(-3{m_{D_i}^2\over\mz^2\cb^2}
    \left[\cww+\sww(e_{U_i}+e_{D_i})\right]
+1+4e_{D_i}\sww+4e_{D_i}^2\sw^4\right)t_Z^{\widetilde Q_i}\right]\right\}
\,.}\eqn\cthree$$
We remind the reader that
$\widetilde\lambda_3\equiv\lambda_3+\lambda_4+\lambda_5$
(in the leading log approximation $\lambda_5 = 0$), the number of
colors $N_{c_i} = 3$ [1] and the electric charges of the quarks
[leptons] are $e_{U_i} = 2/3$ [0] and $e_{D_i} = -1/3$ [-1].
In addition, we have introduced the following notation
$$
\eqalign{
t_Z^X&\equiv\ln\left({M_X^2\over\mzz}\right)\,,\qquad M_X\geq\mz\,,\crr
t_{U_i}^X&\equiv\cases{t_Z^X\,,&$i=1,2$\cr
\ln\left({M_X^2\over m_t^2}\right)\,,&$i=3$}\cr}\eqn\tdefs
$$
where $M_X=|M_1|, |M_2|, |\mu|,  M_{\widetilde Q_i},
M_{\widetilde U_i}, M_{\widetilde D_i}$ for $X=\widetilde B,
\widetilde W, \widetilde H, \widetilde{Q_i}, \widetilde{U_i},
\widetilde{D_i}$.
Furthermore, we have defined $\widetilde\chi_1, \widetilde\chi_2,
\widetilde\chi_{12}$ such that
$$\eqalign{%
M_{\widetilde\chi_1}&= \mu_1 \equiv {\rm max}\{|\mu|, |M_1|\} \cr
M_{\widetilde\chi_2}&= \mu_2 \equiv {\rm max}\{|\mu|, |M_2|\} \cr
M_{\widetilde\chi_{12}}&= \mu_{12}
\equiv {\rm max}\{|\mu|, |M_1|, |M_2|\} \cr
}\eqn\mchis$$
If $M_X<m_Z$ for some field $X$, simply set the
corresponding $t^X_Z=0$.

The couplings $\lambda_1$, $\lambda_2$, and $\widetilde\lambda_3$
evaluated at the scale $\mz$ appear in the CP-even
neutral Higgs mass matrix
and couplings.  For the charged Higgs mass and couplings we require
$\lambda_4$ evaluated at $\mw$.  We find
$$\eqalign{
 \lambda_4&(\mw)=-\half g_2^2(\mw)-{g_2^4\over192\pi^2}
\left\{\vbox to 21pt{}  -6 t_W^{\widetilde\chi_2}+6\tan^2\theta_W
    t_W^{\widetilde\chi_1}+24\tan^2\theta_W
    t_W^{\widetilde\chi_{12}}\right.\cr
   &\left.-4 t_W^{\widetilde H} -8t_W^{\widetilde W}+\sum_i N_{c_i}
\left[6{m_{U_i}^2m_{D_i}^2\over\mw^4\sb^2\cb^2}
-3{m_{U_i}^2\over\mww\sb^2}-3{m_{D_i}^2\over\mww\cb^2}+2\right]
t^{\widetilde Q_i}_{U_i}\right\}\,.
}\eqn\cfour
$$
where the definitions of $t^X_W$ and $t^X_{U_i}$ are obtained from
eq.~\tdefs\ by replacing $m_Z$ with $m_W$.  Finally,
in the leading-log
approximation, $\lambda_i(s)=0$ for $i=5,6$ and 7 at all scales
$\sqrt{s}$.

\Appendix{D}

\medskip
\centerline{\bf One-Loop Squark Contributions to the Scalar Potential}
\medskip

In this appendix we present the derivation of the
Higgs four-point functions from the one-loop effective potential.
The (one generation) squark mass matrix is given by
$$
\calm^2 = \calm_M^2 + \calm_\Gamma^2 + \calm_\Lambda^2\,,
$$
where
$$
\left(\calm_X^2\right)_{ab} \equiv
{\partial^2\calv_X\over\partial\Psi_a\partial\Psi_b^*}\,.\eqn\blabla
$$
for $X = M$, $\Gamma$ and $\Lambda$ and $\calv_X$
given in eq.~\potsqu.
In the special case $M_{\widetilde U}=M_{\widetilde D}=M_{\widetilde Q}
\equiv\msusy$ [\ie, $\calm_M^2 = \one\msusy^2$] we can expand the
effective potential [eq.~\poteff]
$$\eqalign{
\calv &= \calv^0 + {1\over32\pi^2}\tr\left\{\msusy^4
\left[\ln\left({\msusy^2
\over \mu^2}\right)-\threehalf\right]\tr\one\right.\crr
&+2\msusy^2\left[\ln\left({\msusy^2\over \mu^2}\right)-1\right]
\tr\left(M_\Lambda^2+M_\Gamma^2\right)\crr
 &+\ln\left({\msusy^2\over \mu^2}\right)
\tr\left(M_\Lambda^2+M_\Gamma^2\right)^2
+{1\over3\msusy^2}\tr\left(M_\Gamma^2+M_\Lambda^2\right)^3\crr
&\left.-{1\over12\msusy^4}\tr\left(M_\Gamma^2+M_\Lambda^2\right)^4
+\calo\left(\Phi^6\right)\right\}\,.
}\eqn\blabla$$
If we keep in mind that $M_\Gamma$ ($M_\Lambda$) contains one (two)
power(s) of $\Phi$ we find the quartic terms of the potential
$$\eqalign{
\calv_{\rm quartic} &= \Gamma_{ik}^{jl}\left(\Phi^{\dag}_i\Phi_j\right)
\left(\Phi_k^{\dag}\Phi_l\right)+{1\over32\pi^2}
\left\{\ln\left({\msusy^2\over\mu^2}\right)
\tr\left(M_\Lambda^2\right)^2 \right.\cr
&\left.+{1\over\msusy^2}\tr\left(M_\Gamma^2\right)^2M_\Lambda^2
-{1\over12\msusy^4}\tr\left(M_\Gamma^2\right)^4 \right\}
\,.}\eqn\blabla$$
The traces can now be computed without
diagonalization of the mass matrix.
It is now straightforward to absorb the one-loop corrections into the
tree-level terms by redefining the Higgs
tree-level coupling constants as indicated in
eq.~\dlmbdv\ and \dlmbdav.

In the more general case of unequal diagonal squark mass
parameters, the computation of ${\cal V}_{\rm quartic}$ is more
complicated.  Here, one must compute the eigenvalues of the
squark masses (perturbatively in $\mweak^2/\msusy^2$)
before taking the traces.  Details of the more general computation
and the resulting shifts in the Higgs tree-level couplings can be
found in ref.~[\ralfthesis].

\def\refout{\par\penalty-400\vskip\chapterskip
   \spacecheck\referenceminspace
   \ifreferenceopen \Closeout\referencewrite \referenceopenfalse \fi
   \line{\hfil \fourteenrm REFERENCES\hfil}\vskip\headskip
   \input \jobname.refs
   }
\endpage
\refout
\endpage
\figout
\end